\documentclass[10pt,aps,prl,twocolumn,preprintnumbers,nofootinbib]{revtex4-2}
\usepackage{amsmath,amssymb,mathtools}
\usepackage{amsfonts}
\usepackage{hyperref}
\usepackage{xspace}
\usepackage{graphicx}
\usepackage{grffile}
\usepackage{enumitem}
\usepackage{xcolor}
\usepackage{booktabs}
\newcommand{\dd}{\mathrm{d}}

\newcommand{\code}[1]{\texttt{#1}}
\newcommand{\NNLOJET}{\textsc{NNLOjet}\xspace}
\newcommand{\JETSET}{\textsc{JETSET 7.4}\xspace}
\newcommand{\ycut}{y_\mathrm{cut}}
\newcommand{\jet}{\mathrm{jet}}
\newcommand{\njets}{n_\mathrm{jets}}

\newcommand{\NPCnlo}{\code{NPC23\_PI0\_nlo}\xspace}
\newcommand{\NPCnnlo}{\code{NPC23\_PI0\_nnlo}\xspace}
\newcommand{\BDSS}{\code{BDSS21FF\_NLO\_PI0}\xspace}
\newcommand{\BDSSV}{\code{BDSSV22FF\_NNLO\_PI0}\xspace}
\newcommand{\NPCeta}{\code{NPC23\_Eta\_nlo}\xspace}
\newcommand{\ALMSS}{\code{ALMSS25\_ETA\_nlo}\xspace}

\begin{document}

\preprint{CERN-TH-2026-019, ZU-TH 03/26}

\title{\boldmath Precise QCD Predictions for Hadron-in-jet Production in \texorpdfstring{$e^+e^-$}{e+e-} Collisions}

\author{Leonardo Bonino$^a$, Aude Gehrmann-De Ridder$^{a,b}$,\\ Thomas Gehrmann$^a$,
Alexander Huss$^c$, Francesco Merlotti$^b$, Giovanni Stagnitto$^d$}
\affiliation{$^a$
  Physik-Institut, Universit\"at Z\"urich, Winterthurerstrasse 190, 8057 Z\"urich, Switzerland\\
$^b$
   Institute for Theoretical Physics, ETH, CH-8093 Zürich, Switzerland\\
   $^c$  Theoretical Physics Department, CERN, 1211 Geneva 23, Switzerland\\
$^d$  Universit\`{a} di Genova \& INFN, Via Dodecaneso 33, I-16146 Genova, Italy}

\begin{abstract}
The production of identified hadrons inside jets in $e^+ e^-$ annihilation allows for detailed studies of parton-to-hadron fragmentation functions in a clean environment.
We compute hadron-in-jets production cross sections for $e^+e^- \to 2\,\mathrm{jets}$ and $e^+e^- \to 3\,\mathrm{jets}$ to next-to-next-to-leading order (NNLO) in perturbative QCD. By comparing with data from the ALEPH experiment, we demonstrate the implications of the newly computed theory predictions for precision phenomenology.
\end{abstract}

\maketitle

\textit{Introduction---}\ignorespaces
Studying identified hadrons inside jets provides valuable insights into jet formation, its partonic composition, and on the parton-to-hadron fragmentation~\cite{Procura:2009vm,Jain:2011xz,Kaufmann:2015hma}. 
These studies are best performed in the clean environment of $e^+e^-$ annihilation. 
The LEP experiments provided a substantial amount of precision measurements of identified-hadron production cross sections.
These data have been instrumental in the tuning of hadronization models in multi-purpose event simulation programs.
For these data to be equally relevant to precision QCD studies, theory predictions that allow to describe the full kinematics of the final state containing the identified hadron are required.

Precision calculations for identified hadron cross sections have been performed up to now solely for semi-inclusive observables in $e^+e^-$ annihilation (SIA), which are differential only in the energy fraction of the identified hadron in the event, but fully inclusive on the rest of the final state.
For SIA, QCD corrections up to third order in perturbation theory (N\textsuperscript{3}LO) are known~\cite{Altarelli:1979kv,Rijken:1996vr,Rijken:1996ns,Mitov:2006ic,He:2025hin}.
These are routinely included in fits of parton-to-hadron fragmentation functions (FFs) for various hadron species at next-to-leading order (NLO)~\cite{Binnewies:1994ju,Binnewies:1995pt,Kniehl:2000fe,Bourhis:2000gs,Kretzer:2000yf,Kretzer:2001pz,deFlorian:2007ekg,deFlorian:2014xna,deFlorian:2017lwf,Borsa:2021ran,Hirai:2007cx,Albino:2008fy,Bertone:2018ecm,Khalek:2021gxf,Moffat:2021dji,Gao:2024dbv,Gao:2024nkz,Gao:2025bko} and next-to-next-to-leading order (NNLO)~\cite{Bertone:2017tyb,Anderle:2015lqa,Abdolmaleki:2021yjf,Borsa:2022vvp,Gao:2025hlm,AbdulKhalek:2022laj,Li:2024etc}.
These fits usually combine data on SIA, semi-inclusive deep inelastic scattering (SIDIS), also known to NNLO~\cite{Altarelli:1979kv,deFlorian:1997zj,Bonino:2024qbh,Goyal:2023zdi}, and identified hadron production at hadron colliders.
Semi-inclusive coefficient functions for the latter are known to NLO~\cite{Aversa:1988vb,Jager:2002xm}, and fully differential cross sections at NNLO have been obtained most recently~\cite{Czakon:2025yti,Generet:2025bqx}.

Future electron--positron colliders~\cite{FCC:2025lpp} will produce enormous data sets of hadronic final states, thereby allowing a new era of precision QCD studies~\cite{Zhou:2024cyk,dEnterria:2025pml}.
To enable these studies, theory predictions must go beyond the simple SIA kinematics and include full final state descriptions for identified hadron observables.
In this letter, we present the first NNLO QCD calculation for identified hadron observables in 2-jet and 3-jet final states in $e^+e^-$ annihilation.
We demonstrate the impact of these corrections through a precise phenomenological study of ALEPH data on neutral light mesons in jet final states, whilst comparing predictions obtained employing recent determinations of FFs.
\newline

\textit{Methodology---}\ignorespaces
We consider the production of 2-jet and 3-jet final states at $e^+e^-$ colliders, with one of the jets containing
an identified hadron $H$:
\begin{align}\label{eq:proc}
  e^+e^-& \to \jet(H)+\jet
  \, , \nonumber \\
  e^+e^-& \to \jet(H)+\jet + \jet
  \, .
\end{align}
In the fully differential cross section for these processes,
\begin{align}\label{eq:xs}
\dd\sigma^H=\sum_p\dd\sigma^{p\to H} = \sum_p \int\dd z D^H_p(z,\mu_a^2)\dd \hat{\sigma}_p(z,\mu_a^2) \, ,
\end{align}
the fragmentation function $D_p^H$ encodes the probability for the transition of the parton $p$ into the identified hadron $H$.
Collinear factorization is performed at a fragmentation scale $\mu_a$.
The short-distance one-parton-identified cross section $\dd \hat{\sigma}_p$ admits a perturbative expansion in the renormalized strong coupling constant $\alpha_s$
\begin{align}
\dd \hat{\sigma}_p = \dd \hat{\sigma}^{\mathrm{LO}}_p+\left(\frac{\alpha_s}{2\pi}\right) \dd \hat{\sigma}^{\mathrm{NLO}}_p+\left(\frac{\alpha_s}{2\pi}\right)^2 \dd \hat{\sigma}^{\mathrm{NNLO}}_p+\ldots \, .
\end{align}
Beyond leading order (LO), individual real and virtual contributions to the short distance
cross section contain infrared divergences.
To deal with such divergences at intermediate steps of the calculation, we adopt the \textit{antenna subtraction method}~\cite{Gehrmann-DeRidder:2005btv,Currie:2013vh}.
In this method, with the introduction of subtraction terms, infrared divergences associated to single- and double-unresolved configurations in real radiation matrix elements are extracted and recombined with virtual and collinear factorization contributions.
The building blocks of the subtraction terms are antenna functions that encode all possible radiation patterns between a pair of hard radiating partons.

Since the fragmentation process is sensitive to the momentum fraction of individual partons within a collinear cluster, the identification of a parton in \eqref{eq:xs} affects the cancellation of collinear divergences.
The antenna subtraction method has been extended to account for identified hadrons as described in~\cite{Gehrmann:2022cih,Gehrmann:2022pzd,Bonino:2024adk}. 
This extended formalism was first employed in the computation of charged hadrons inside jets in association with a leptonically decaying $Z$ boson at the LHC~\cite{Caletti:2024xaw}.

In this work, we implement identified-hadron production for $e^+e^- \to 2\,\mathrm{jets}$ and $e^+e^- \to 3\,\mathrm{jets}$ processes through NNLO accuracy within the parton-level event generator \NNLOJET~\cite{NNLOJET:2025rno}.
The subtraction terms for these processes structurally resemble the ones for jet production in deep-inelastic scattering (DIS)~\cite{Currie:2017tpe}: initial-state collinear singularities in DIS are related to final-state collinear singularities in identified hadron production through crossing symmetry of matrix elements and subtraction terms.

Identified hadron production in $e^+e^- \to 2\,\mathrm{jets}$ is validated against an implementation~\cite{Bonino:2023icn} of the NNLO coefficient functions~\cite{Rijken:1996vr,Rijken:1996ns,Mitov:2006ic} for SIA, where the convolution of coefficient functions and fragmentation functions is performed analytically in Mellin space.
\newline

\textit{Phenomenological Setup---}\ignorespaces
The bulk of LEP data with identified hadrons in the final state are for SIA reactions, whereas only a limited set of measurements are available for hadron-in-jet production~\cite{ALEPH:1999udi,L3:1995rnv,OPAL:2004prv}.

The ALEPH collaboration measured the production of identified neutral light hadrons ($\pi^0$ and $\eta$ mesons) in jet events~\cite{ALEPH:1999udi}, based on data collected at the $Z$ peak ($\sqrt{s}= M_Z = 91.2$ GeV).
Measured cross sections are presented differential in the observable
\begin{align}\label{eq:x}
x = \frac{2 E_H}{\sqrt{s}} \, ,
\end{align}
where $E_H$ is the energy of the identified hadron. 
The differential cross sections are normalized to the fiducial cross section $\sigma_{\njets}$ for the production of $\njets$ jets.

Jets are reconstructed from event tracks using the Durham jet algorithm~\cite{Catani:1991hj} with $\ycut = 0.01$.
In the study on hadron production in 3-jet events, jets are ordered according to their energy, providing distributions in $x$ for the leading (LJ), sub-leading (SLJ), and sub-sub-leading jet (SSLJ).
Events are accepted for $\theta_T\in \left[30^\circ,150^\circ\right]$ where $\theta_T$ is the polar angle between the thrust direction and the beam axis.
Jets are required to have a polar angle between the jet and the beam axis $\theta_j \in \left[30^\circ,150^\circ\right]$.
These criteria define fiducial cross sections for a given jet multiplicity. 
In their analysis~\cite{ALEPH:1999udi}, the ALEPH collaboration corrected the data to full $4\pi$ solid angle acceptance using parton-shower event simulations.

To demonstrate the feasibility of a data--theory comparison at fiducial cross section level, we recompute the
acceptance correction factors with \JETSET~\cite{Sjostrand:1994kzr} in analogy to the prescription in the ALEPH analysis.
We then use these correction factors to revert the data to fiducial cross section level.
Correction factors are between $0.95$ and $1.02$ for 2-jet $\pi^0$ data, $0.93$ and $0.99$ for 2-jet $\eta$ data, and between $0.96$ and $1.00$ for 3-jet $\pi^0$ data. In general, they exhibit only a moderate $x$ dependence.
In the phenomenological analysis that follows, all data are corrected with these acceptance factors.

We present the first NNLO QCD predictions for the observables measured in~\cite{ALEPH:1999udi}.
The choice of renormalization and fragmentation scales deserve particular attention.
For the renormalization scale $\mu_r$ we adopt $\mu_r = \sqrt{s}$, as the hard scale of the overall process, and use $\alpha_s(M_Z)=0.118$.
Regarding the fragmentation scale $\mu_a$, in semi-inclusive calculations the only natural choice is $\sqrt{s}$.
In contrast, when hadrons are identified within jets, the maximum size of the jet $\ycut$ introduces an intrinsic cut-off for extra radiation, thereby reducing the fragmentation scale~\cite{Dokshitzer:1987nm,Dokshitzer:1991wu,Procura:2009vm,Jain:2011xz,Kaufmann:2015hma}.
We shall examine different choices of fragmentation scales in our QCD predictions.

We investigate the impact of the FF sets used in predictions for 2-jet and 3-jet events for $\pi^0$ production by comparing four different FF sets from the BDSSV and the NPC collaborations: \BDSS~\cite{Borsa:2021ran}, \BDSSV~\cite{Borsa:2022vvp}, \NPCnlo~\cite{Gao:2025bko}, and \NPCnnlo~\cite{Gao:2025hlm}.
Both collaborations provide a NNLO set fitted to SIA and SIDIS data, and a NLO set with an improved description of $D_g^{\pi^0}$ due to the inclusion of proton--proton collision data in the fit.
For 2-jet events, we also focus on $\eta$ meson production and compared the results using the NLO FF sets \ALMSS~\cite{Aidala:2025kep} and \NPCeta~\cite{Gao:2025bko}.
\newline

\begin{figure}[ht]
\centering
\includegraphics[width=\columnwidth]{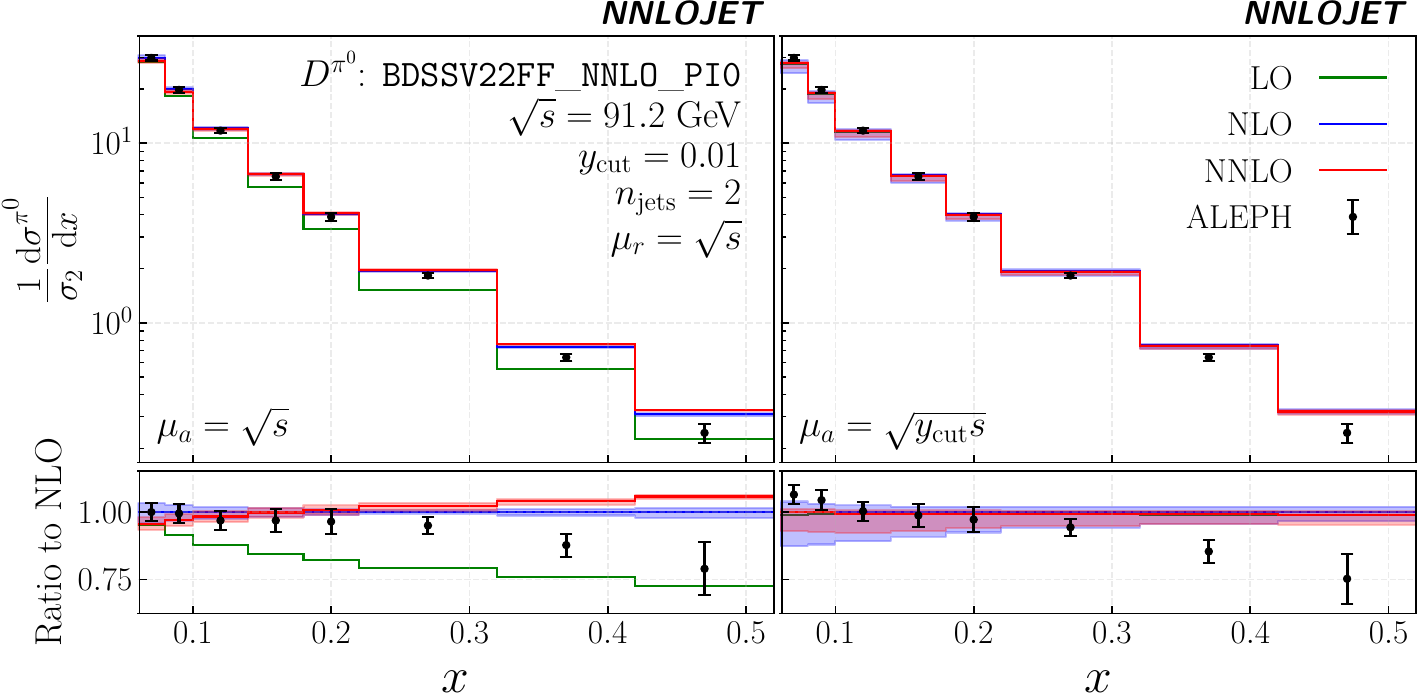}
\caption{Data--theory comparison for $\pi^0$ production in 2-jet exclusive events up NNLO with \BDSSV. Left panel: $\mu_a = \sqrt{s}$. Right panel: $\mu_a = \sqrt{\ycut s}$.}
\label{fig:2j_pi0}
\end{figure}
\textit{2-jet Results---}\ignorespaces
In Fig.~\ref{fig:2j_pi0} we compare our theoretical calculation to ALEPH data for $\pi^0$ production in 2-jet exclusive events.
Predictions up to NNLO are computed with the NNLO FF set \BDSSV.
Higher-order theory uncertainties, here and in what follows, are estimated with a 7-point scale variation of $\mu_r$ and $\mu_a$ around the central scales.
We investigate the effect of the fragmentation scale choice by comparing our predictions for two values of the central scale, $\mu_a=\sqrt{s}$ and $\mu_a=\sqrt{\ycut s}$.
We observe that the perturbative convergence of the calculation drastically improves for $\mu_a=\sqrt{\ycut s}$, with reduced perturbative corrections and good agreement between theory and data.
This observation confirms the assertion~\cite{Dokshitzer:1987nm,Dokshitzer:1991wu,Procura:2009vm,Jain:2011xz,Kaufmann:2015hma} that the typical hadron-in-jet production scale is determined by the individual jet kinematics rather than the full event kinematics.

\begin{figure}[ht]
  \centering
  \includegraphics[width=\columnwidth]{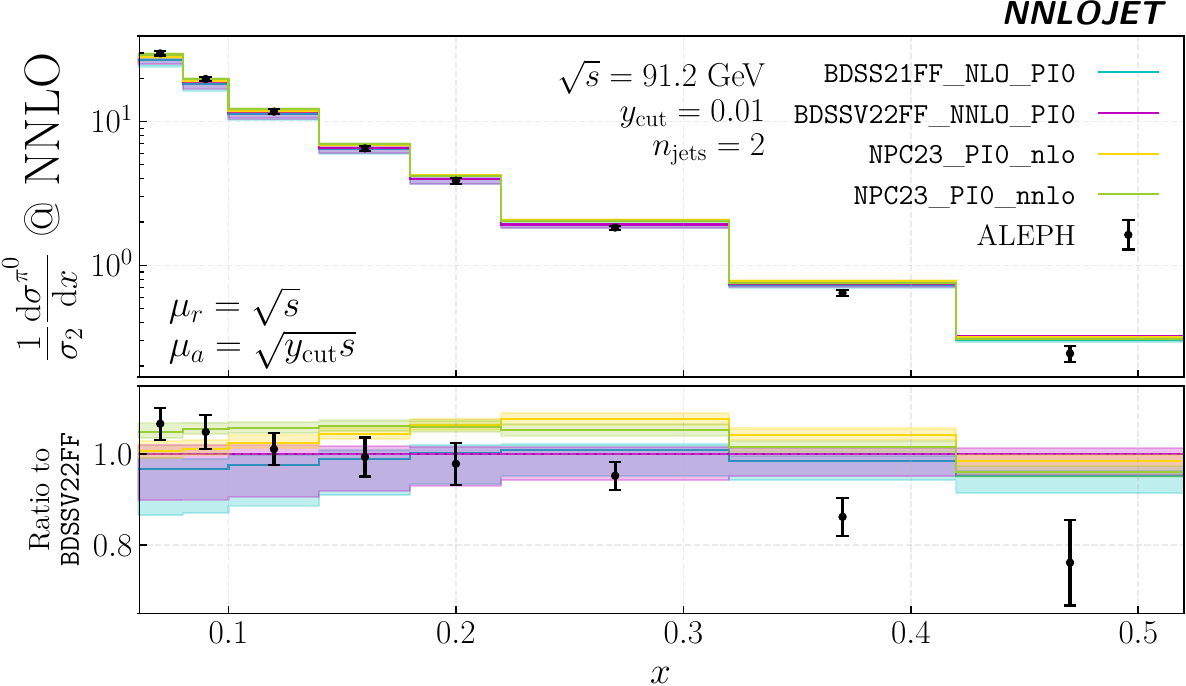}
  \caption{Data--theory comparison of $\pi^0$ production in 2-jet exclusive events at NNLO with different FF sets.
  }
  \label{fig:2j_pi0_sets}
\end{figure}

In Fig.~\ref{fig:2j_pi0_sets} we assess the performance of the NNLO predictions of $\pi^0$ 2-jet events computed with the selected FF sets.
We observe that the predictions are compatible within theoretical uncertainties for a given collaboration in the entire $x$ range, while for $x<0.32$ the predictions from BDSSV and NPC do not overlap.
The BDSSV FFs provide a better quantitative description of the data.

We find that the  dominant source of theoretical uncertainty comes from the variation of the fragmentation scale. 
As previously observed in SIA~\cite{Rijken:1996vr,Rijken:1996ns}, the fragmentation-scale dependence of quark- and gluon-initiated contributions exhibits a partial cancellation at higher perturbative orders.
For $z<0.1$ at $\mu_a=\sqrt{\ycut s}$, the  
BDSSV sets differ considerably from the NPC sets. The sea-like shape of BDSSV  in this range 
results in considerably larger absolute values of the quark and gluon induced contributions than the valence-like shape of NPC, leading to  
larger scale-variation uncertainty bands in the differential distributions.
\begin{figure}[ht]
\centering
\includegraphics[width=\columnwidth]{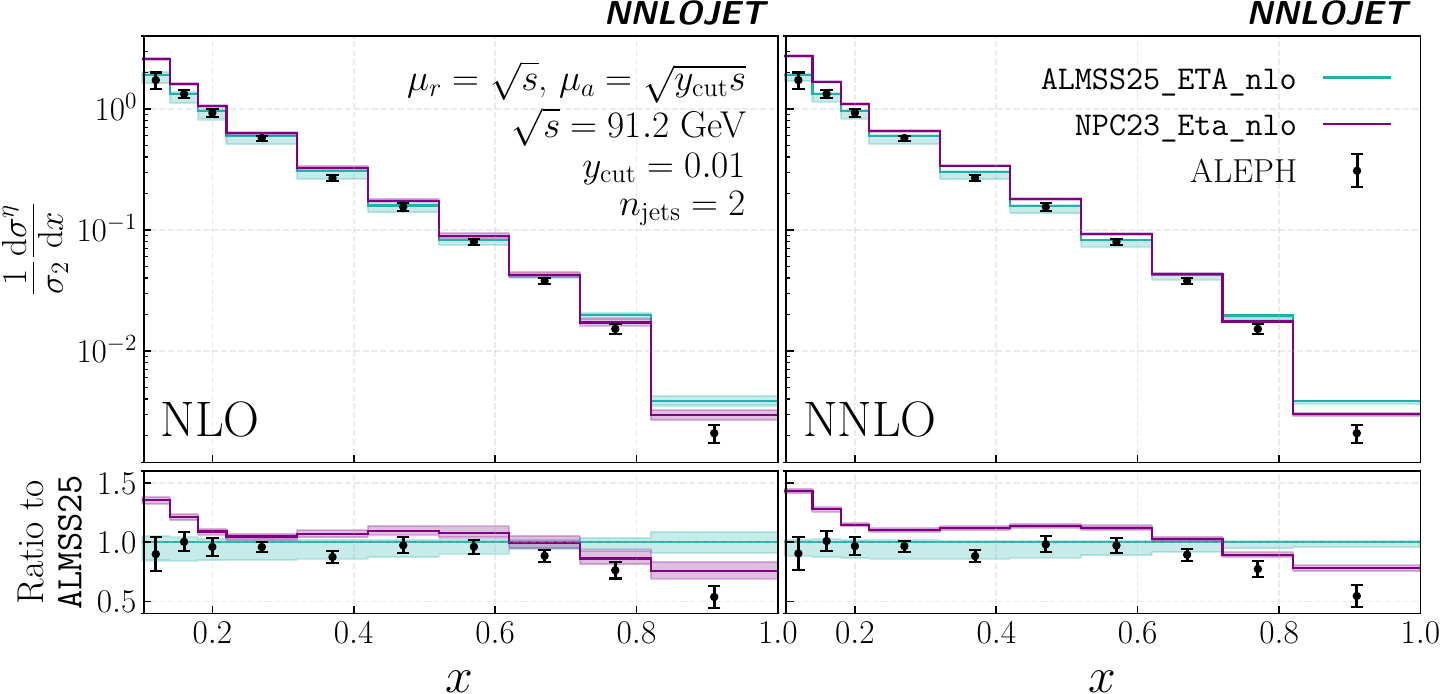}
\caption{Data--theory comparison for $\eta$ production in 2-jet exclusive events up NNLO for \ALMSS (green) and \NPCeta (purple) FF sets. Left panel: NLO. Right panel: NNLO.
}
\label{fig:2j_eta}
\end{figure}

In Fig.~\ref{fig:2j_eta} we compare our theoretical predictions up to NNLO against ALEPH data for $\eta$ meson production in 2-jet exclusive events using the NLO FF sets \ALMSS and \NPCeta.
The $\eta$ data cover a larger kinematic range in $x$ as compared to the $\pi^0$ data.
Both sets provide a satisfactory description of the data for $0.2<x<0.7$.
While the \ALMSS provides a better description of the data at small $x$, the \NPCeta performs better at large $x$.
Overall, predictions computed with the \ALMSS FF set describe the data the best.

As observed for $\pi^0$ 2-jet predictions, theoretical uncertainties are dominated by the fragmentation scale variation.
Predictions obtained with the \ALMSS set are more sensitive to $\mu_a$ variations as compared to the ones produced with \NPCeta.

Accounting for FF uncertainties does not change the conclusions drawn above on the description of the data for both $\pi^0$ and $\eta$ in 2-jet observables.
Lastly, we note that none of the FF fits include the $\pi^0$ and $\eta$ 2-jet exclusive data.
\newline

\begin{figure*}[t]
\centering
\includegraphics[width=\textwidth]{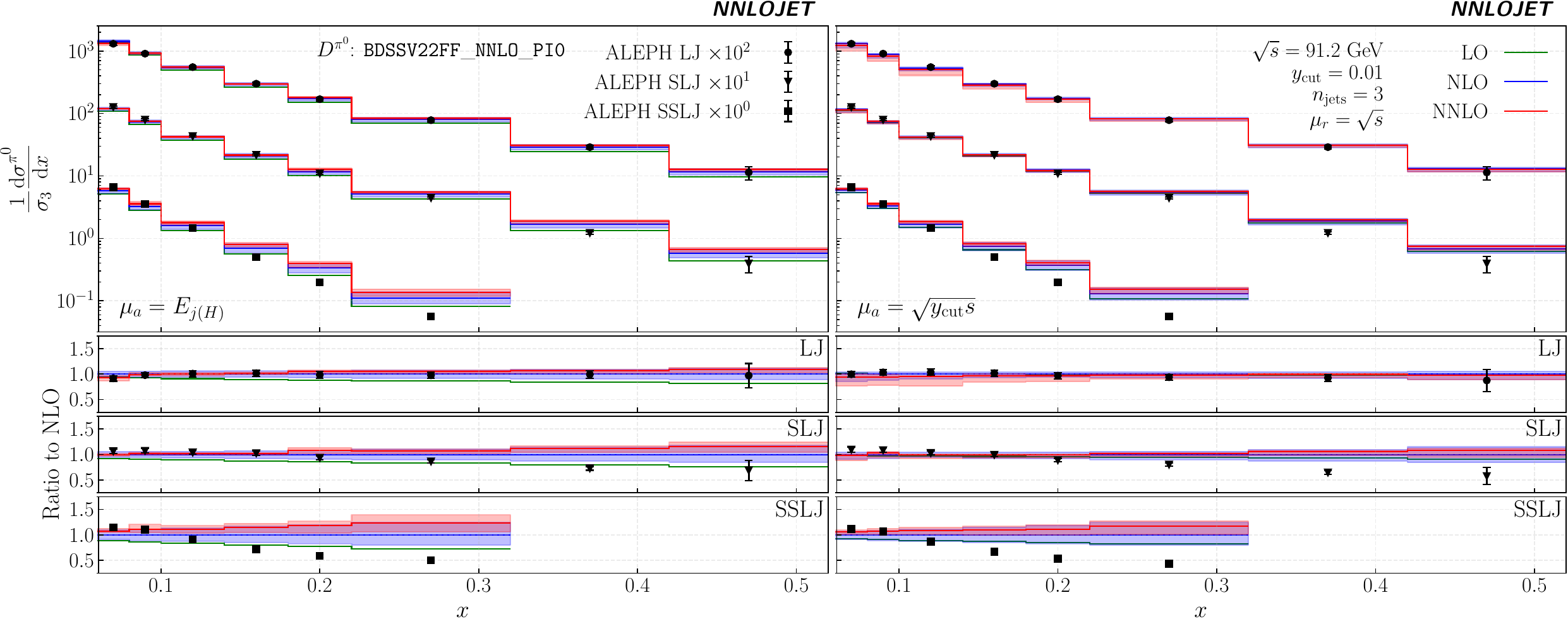}
\caption{Data--theory comparison for $\pi^0$ production in 3-jet exclusive events up NNLO with \BDSSV. Left panel: $\mu_a = E_{j(H)}$. Right panel: $\mu_a = \sqrt{\ycut s}$.
}
\label{fig:3j_pi0_scales}
\end{figure*}

\textit{3-jet Results---}\ignorespaces
Identified hadron production in 3-jet final states is particularly relevant due to its direct sensitivity
to the gluon-to-hadron FF.

Similarly to the 2-jet observables, we investigate the effect of the choice of the fragmentation scale.
To produce predictions for 3-jet final states, we can choose between a jet resolution based scale $\mu_a=\sqrt{\ycut s}$ and a jet energy based scale $\mu_a=E_{j(H)}$ corresponding to the energy of the jet containing the identified hadron.
In Fig.~\ref{fig:3j_pi0_scales} we compare our predictions up to NNLO computed with the \BDSSV FF set for the two choices of the fragmentation scale against $\pi^0$ 3-jet data.
Both fragmentation scale choices provide comparable predictions. 
They yield an excellent description of the LJ data, which degrades consecutively for SLJ and SSLJ. 
Predictions for both choices show good perturbative convergence, with $\mu_a=\sqrt{\ycut s}$ featuring slightly smaller corrections and lower scale uncertainties.

\begin{figure}[t]
\centering
\includegraphics[width=\columnwidth]{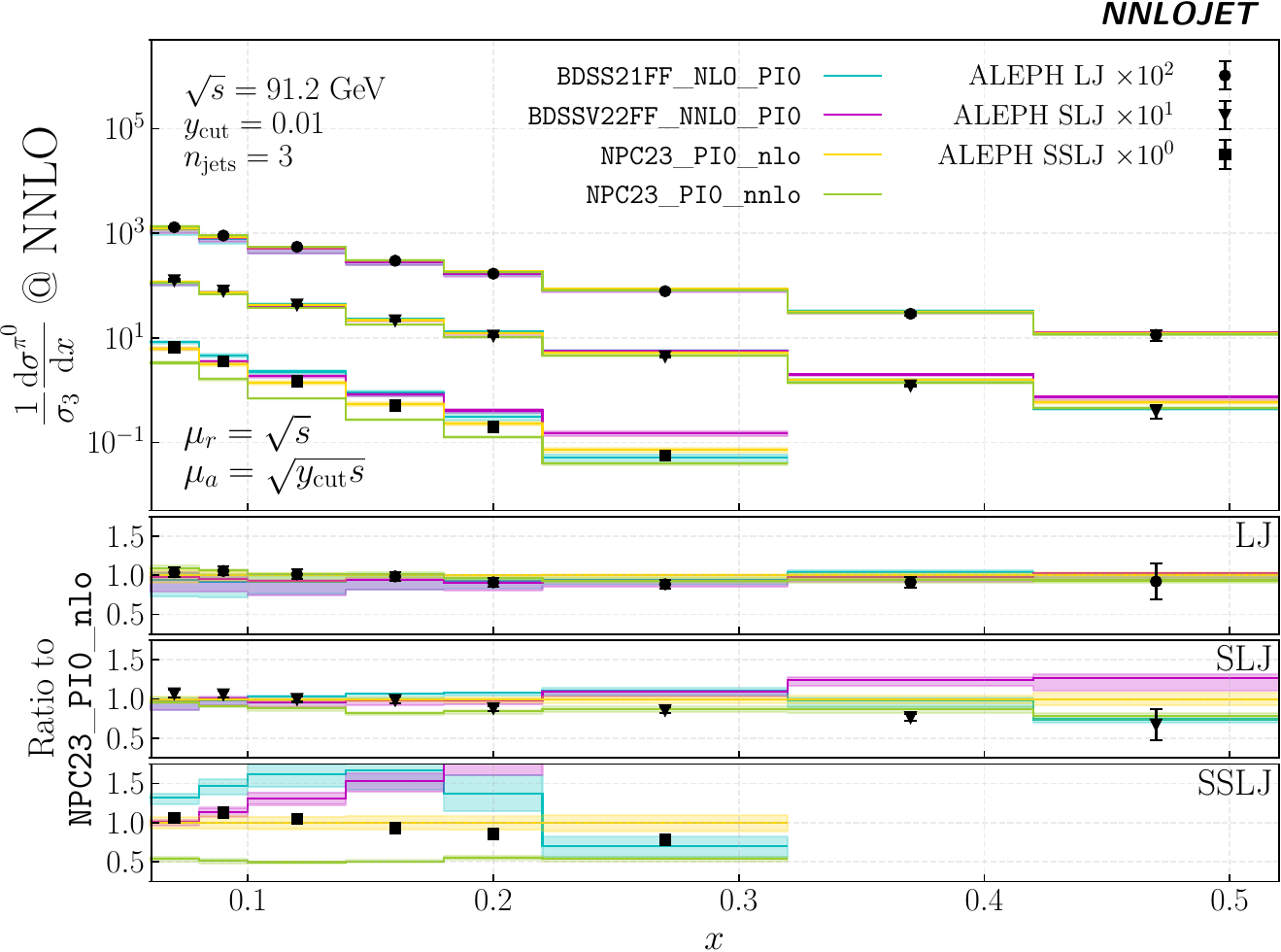}
\caption{Data--theory comparison of $\pi^0$ production in 3-jet exclusive events at NNLO with different FF sets.
}
\label{fig:3j_pi0_sets}
\end{figure}

The \BDSSV set used in Fig.~\ref{fig:3j_pi0_scales} has been fitted only to SIA and SIDIS data, which are both largely dominated by quark fragmentation processes. 
The resulting uncertainty on the gluon FF is the most likely explanation for the degrading description of the distributions in the SLJ and SSLJ.
To gain further insights into this matter, in Fig.~\ref{fig:3j_pi0_sets} we compare the NNLO predictions (for $\mu_a=\sqrt{\ycut s}$) for the selected FF sets.
The LJ predominantly originates from quarks, and predictions for the LJ distribution with all FF sets describe the data equally well, reflecting the accuracy on the $D_q^{\pi^0}$ FFs.
The impact of gluon-initiated processes increases from LJ to SLJ and SSLJ.
Only the \BDSS and \NPCnlo sets include proton--proton collision data in their global fits, thereby providing more reliable constraints on the $D_g^{\pi^0}$ FF, which is reflected in better descriptions of the ALEPH data for the SLJ and SSLJ.
In particular the \NPCnlo set, which uses this specific dataset as a test observable, shows the best agreement with the data.
The improvement in the description of SLJ and SSLJ data confirms the relevance of the ALEPH data in constraining the gluon FF.

As in the 2-jet analysis, the uncertainty from scale variations is predominantly driven by the fragmentation scale.
The larger overall uncertainty bands obtained with the BDSSV sets can be traced back to the small-$z$ behaviour of the corresponding fragmentation functions.
Unlike the 2-jet case, where the contribution of $D_g^{\pi^0}$ only plays a marginal role, the 3-jet observables show a stronger dependence on the FF uncertainties.
Nevertheless, these FF uncertainties do not alter the overall pattern of agreement among the different sets displayed in Fig.~\ref{fig:3j_pi0_sets}.
\newline

\textit{On Gluon Fragmentation---}\ignorespaces
To illustrate the role of the gluon FF we perform a breakdown of the total fiducial 3-jet cross section into the $q$- and $g$-initiated fragmentation processes.
To this end, we define
\begin{align}
  R^{p \to \pi^0} = \frac{\sigma_3^{p \to \pi^0}}{\sigma_3^{\pi^0}}\, ,
\end{align}
as the ratio of fiducial cross sections for $\pi^0$ production in 3-jet events initiated through parton-type $p$ divided by the sum of all parton types.
In addition, to further investigate $\pi^0$ production separately for each jet of the 3-jet event, we introduce
\begin{align}
  R_{\jet}^{p \to \pi^0} = \frac{\sigma_{\jet}^{p \to \pi^0}}{\sigma_{\jet}^{\pi^0}}\, ,
\end{align}
where $\jet$ denotes the identification of the $\pi^0$ meson within the LJ, SLJ, or SSLJ.

\renewcommand{\arraystretch}{1.25}
\begin{table}[ht]
  \centering
  \begin{tabular}{p{1.2cm}rrr}
    \toprule
    \toprule
    & \textbf{LO} & \textbf{NLO} & \textbf{NNLO} \\
    \midrule
    $R_{\rm LJ}^{q \to \pi^0}$ & 95.385\% & 96.477\% & 100.50\% \\
    $R_{\rm LJ}^{g \to \pi^0}$ &  4.615\% &  3.523\% &  $-0.50$\% \\
    \midrule
    $R_{\rm SLJ}^{q \to \pi^0}$ & 81.027\% & 78.86\% & 79.4\% \\
    $R_{\rm SLJ}^{g \to \pi^0}$ & 18.973\% & 21.14\% & 20.6\% \\
    \midrule
    $R_{\rm SSLJ}^{q \to \pi^0}$ & 46.644\% & 40.45\% & 35.8\% \\
    $R_{\rm SSLJ}^{g \to \pi^0}$ & 53.356\% & 59.55\% & 64.2\% \\
    \midrule
    \midrule
    $R^{q \to \pi^0}$ & 84.161\% & 82.480\% & 83.26\% \\
    $R^{g \to \pi^0}$ & 15.839\% & 17.520\% & 16.74\% \\
    \bottomrule
    \bottomrule
  \end{tabular}
  \caption{Quark- and gluon-initiated fractions of the fiducial cross sections for identified hadron production in 3-jet events up to NNLO separating LJ, SLJ, and SSLJ contributions. Predictions are computed with $\mu_a = \sqrt{\ycut s}$ and the \NPCnlo FF set.}
  \label{tab:id_parton_3jet}
\end{table}
\renewcommand{\arraystretch}{1}

In Tab.~\ref{tab:id_parton_3jet} we report the values of $R^{p \to \pi^0}$ and $R_{\jet}^{p \to \pi^0}$ for $p=q,g$ up to NNLO for the \NPCnlo set and $\mu_a=\sqrt{\ycut s}$.
The fractions change only moderately with increasing perturbative orders, except for the SSLJ.
$\pi^0$ mesons in the LJ originate almost entirely from the fragmentation of a quark.
The contribution from gluons increases in the SLJ, with $g$-originated fragmentation accounting for 21\% of the events.
In the SSLJ, gluon fragmentation dominates the production of pions.
In this case, substantial higher-order corrections are observed, which increase the gluon-induced fraction from 53\% at LO to 64\% at NNLO.
\newline

\textit{Conclusions---}\ignorespaces
In this letter, we compute the NNLO QCD corrections to identified hadron-in-jet production observables in 2-jet and 3-jet final states in $e^+e^-$ annihilation.
Our novel results establish the perturbative stability of the predictions, provided that the choice of the fragmentation scale takes into account the jet resolution, as advocated in~\cite{Dokshitzer:1987nm,Dokshitzer:1991wu,Procura:2009vm,Jain:2011xz,Kaufmann:2015hma}.

Our calculation is performed in the \NNLOJET parton-level event generator framework, which allows to compute any infrared-safe observable derived from these final states, and to account for geometrical acceptance cuts that define the fiducial cross sections.
It shows that modern-day theory predictions no longer require ad hoc corrections to full geometrical acceptance for precision physics studies on $e^+e^-$ data.

Our results demonstrate the increased sensitivity of the distribution of hadrons in the subleading-energy jets to the poorly constrained gluon-to-hadron fragmentation functions and show the impact of ALEPH hadron-in-jet data in their determination.
Our newly derived NNLO QCD corrections lay the ground for precision QCD studies on hadron-in-jet observables through the reanalysis of LEP data, following recent works on jets~\cite{Electron-PositronAlliance:2021kig} and event shapes~\cite{Electron-PositronAlliance:2025hze}, and at future high-energy $e^+e^-$ colliders.
\newline

\textit{Acknowledgments---}\ignorespaces
We are thankful to Ignacio Borsa and Jun Gao for providing us with the FF sets used in this work.
This work has been supported by the Swiss National Science Foundation (SNF) under contracts 200021-231259 and 240015, by the Swiss National Supercomputing Centre (CSCS) under project IDs ETH5f and UZH10, by the European Union's Horizon 2020 research and innovation programme grant agreement 101019620 (ERC Advanced Grant TOPUP) as well as by the EU Horizon Europe research and innovation programme under the Marie-Sk\l{}odowska Curie Action HIPFLAPP, grant agreement 101149076.

\textit{Data availability---}\ignorespaces
The data that support the findings of this article are openly available \cite{bonino_2026_18683133}.

\bibliography{epem_hfrag}

\begin{thebibliography}{63}%
\makeatletter
\providecommand \@ifxundefined [1]{%
 \@ifx{#1\undefined}
}%
\providecommand \@ifnum [1]{%
 \ifnum #1\expandafter \@firstoftwo
 \else \expandafter \@secondoftwo
 \fi
}%
\providecommand \@ifx [1]{%
 \ifx #1\expandafter \@firstoftwo
 \else \expandafter \@secondoftwo
 \fi
}%
\providecommand \natexlab [1]{#1}%
\providecommand \enquote  [1]{``#1''}%
\providecommand \bibnamefont  [1]{#1}%
\providecommand \bibfnamefont [1]{#1}%
\providecommand \citenamefont [1]{#1}%
\providecommand \href@noop [0]{\@secondoftwo}%
\providecommand \href [0]{\begingroup \@sanitize@url \@href}%
\providecommand \@href[1]{\@@startlink{#1}\@@href}%
\providecommand \@@href[1]{\endgroup#1\@@endlink}%
\providecommand \@sanitize@url [0]{\catcode `\\12\catcode `\$12\catcode
  `\&12\catcode `\#12\catcode `\^12\catcode `\_12\catcode `\%12\relax}%
\providecommand \@@startlink[1]{}%
\providecommand \@@endlink[0]{}%
\providecommand \url  [0]{\begingroup\@sanitize@url \@url }%
\providecommand \@url [1]{\endgroup\@href {#1}{\urlprefix }}%
\providecommand \urlprefix  [0]{URL }%
\providecommand \Eprint [0]{\href }%
\providecommand \doibase [0]{http://dx.doi.org/}%
\providecommand \selectlanguage [0]{\@gobble}%
\providecommand \bibinfo  [0]{\@secondoftwo}%
\providecommand \bibfield  [0]{\@secondoftwo}%
\providecommand \translation [1]{[#1]}%
\providecommand \BibitemOpen [0]{}%
\providecommand \bibitemStop [0]{}%
\providecommand \bibitemNoStop [0]{.\EOS\space}%
\providecommand \EOS [0]{\spacefactor3000\relax}%
\providecommand \BibitemShut  [1]{\csname bibitem#1\endcsname}%
\let\auto@bib@innerbib\@empty
\bibitem [{\citenamefont {Procura}\ and\ \citenamefont
  {Stewart}(2010)}]{Procura:2009vm}%
  \BibitemOpen
  \bibfield  {author} {\bibinfo {author} {\bibfnamefont {M.}~\bibnamefont
  {Procura}}\ and\ \bibinfo {author} {\bibfnamefont {I.~W.}\ \bibnamefont
  {Stewart}},\ }\href {\doibase 10.1103/PhysRevD.81.074009} {\bibfield
  {journal} {\bibinfo  {journal} {Phys. Rev. D}\ }\textbf {\bibinfo {volume}
  {81}},\ \bibinfo {pages} {074009} (\bibinfo {year} {2010})},\ \bibinfo {note}
  {[Erratum: Phys.Rev.D 83, 039902 (2011)]},\ \Eprint
  {http://arxiv.org/abs/0911.4980} {arXiv:0911.4980 [hep-ph]} \BibitemShut
  {NoStop}%
\bibitem [{\citenamefont {Jain}\ \emph {et~al.}(2011)\citenamefont {Jain},
  \citenamefont {Procura},\ and\ \citenamefont {Waalewijn}}]{Jain:2011xz}%
  \BibitemOpen
  \bibfield  {author} {\bibinfo {author} {\bibfnamefont {A.}~\bibnamefont
  {Jain}}, \bibinfo {author} {\bibfnamefont {M.}~\bibnamefont {Procura}}, \
  and\ \bibinfo {author} {\bibfnamefont {W.~J.}\ \bibnamefont {Waalewijn}},\
  }\href {\doibase 10.1007/JHEP05(2011)035} {\bibfield  {journal} {\bibinfo
  {journal} {JHEP}\ }\textbf {\bibinfo {volume} {05}},\ \bibinfo {pages} {035}
  (\bibinfo {year} {2011})},\ \Eprint {http://arxiv.org/abs/1101.4953}
  {arXiv:1101.4953 [hep-ph]} \BibitemShut {NoStop}%
\bibitem [{\citenamefont {Kaufmann}\ \emph {et~al.}(2015)\citenamefont
  {Kaufmann}, \citenamefont {Mukherjee},\ and\ \citenamefont
  {Vogelsang}}]{Kaufmann:2015hma}%
  \BibitemOpen
  \bibfield  {author} {\bibinfo {author} {\bibfnamefont {T.}~\bibnamefont
  {Kaufmann}}, \bibinfo {author} {\bibfnamefont {A.}~\bibnamefont {Mukherjee}},
  \ and\ \bibinfo {author} {\bibfnamefont {W.}~\bibnamefont {Vogelsang}},\
  }\href {\doibase 10.1103/PhysRevD.92.054015} {\bibfield  {journal} {\bibinfo
  {journal} {Phys. Rev. D}\ }\textbf {\bibinfo {volume} {92}},\ \bibinfo
  {pages} {054015} (\bibinfo {year} {2015})},\ \bibinfo {note} {[Erratum:
  Phys.Rev.D 101, 079901 (2020)]},\ \Eprint {http://arxiv.org/abs/1506.01415}
  {arXiv:1506.01415 [hep-ph]} \BibitemShut {NoStop}%
\bibitem [{\citenamefont {Altarelli}\ \emph {et~al.}(1979)\citenamefont
  {Altarelli}, \citenamefont {Ellis}, \citenamefont {Martinelli},\ and\
  \citenamefont {Pi}}]{Altarelli:1979kv}%
  \BibitemOpen
  \bibfield  {author} {\bibinfo {author} {\bibfnamefont {G.}~\bibnamefont
  {Altarelli}}, \bibinfo {author} {\bibfnamefont {R.~K.}\ \bibnamefont
  {Ellis}}, \bibinfo {author} {\bibfnamefont {G.}~\bibnamefont {Martinelli}}, \
  and\ \bibinfo {author} {\bibfnamefont {S.-Y.}\ \bibnamefont {Pi}},\ }\href
  {\doibase 10.1016/0550-3213(79)90062-2} {\bibfield  {journal} {\bibinfo
  {journal} {Nucl. Phys. B}\ }\textbf {\bibinfo {volume} {160}},\ \bibinfo
  {pages} {301} (\bibinfo {year} {1979})}\BibitemShut {NoStop}%
\bibitem [{\citenamefont {Rijken}\ and\ \citenamefont {van
  Neerven}(1996)}]{Rijken:1996vr}%
  \BibitemOpen
  \bibfield  {author} {\bibinfo {author} {\bibfnamefont {P.~J.}\ \bibnamefont
  {Rijken}}\ and\ \bibinfo {author} {\bibfnamefont {W.~L.}\ \bibnamefont {van
  Neerven}},\ }\href {\doibase 10.1016/0370-2693(96)00898-2} {\bibfield
  {journal} {\bibinfo  {journal} {Phys. Lett. B}\ }\textbf {\bibinfo {volume}
  {386}},\ \bibinfo {pages} {422} (\bibinfo {year} {1996})},\ \Eprint
  {http://arxiv.org/abs/hep-ph/9604436} {arXiv:hep-ph/9604436} \BibitemShut
  {NoStop}%
\bibitem [{\citenamefont {Rijken}\ and\ \citenamefont {van
  Neerven}(1997)}]{Rijken:1996ns}%
  \BibitemOpen
  \bibfield  {author} {\bibinfo {author} {\bibfnamefont {P.~J.}\ \bibnamefont
  {Rijken}}\ and\ \bibinfo {author} {\bibfnamefont {W.~L.}\ \bibnamefont {van
  Neerven}},\ }\href {\doibase 10.1016/S0550-3213(96)00669-4} {\bibfield
  {journal} {\bibinfo  {journal} {Nucl. Phys. B}\ }\textbf {\bibinfo {volume}
  {487}},\ \bibinfo {pages} {233} (\bibinfo {year} {1997})},\ \Eprint
  {http://arxiv.org/abs/hep-ph/9609377} {arXiv:hep-ph/9609377} \BibitemShut
  {NoStop}%
\bibitem [{\citenamefont {Mitov}\ \emph {et~al.}(2006)\citenamefont {Mitov},
  \citenamefont {Moch},\ and\ \citenamefont {Vogt}}]{Mitov:2006ic}%
  \BibitemOpen
  \bibfield  {author} {\bibinfo {author} {\bibfnamefont {A.}~\bibnamefont
  {Mitov}}, \bibinfo {author} {\bibfnamefont {S.}~\bibnamefont {Moch}}, \ and\
  \bibinfo {author} {\bibfnamefont {A.}~\bibnamefont {Vogt}},\ }\href {\doibase
  10.1016/j.physletb.2006.05.005} {\bibfield  {journal} {\bibinfo  {journal}
  {Phys. Lett. B}\ }\textbf {\bibinfo {volume} {638}},\ \bibinfo {pages} {61}
  (\bibinfo {year} {2006})},\ \Eprint {http://arxiv.org/abs/hep-ph/0604053}
  {arXiv:hep-ph/0604053} \BibitemShut {NoStop}%
\bibitem [{\citenamefont {He}\ \emph {et~al.}(2025)\citenamefont {He},
  \citenamefont {Xing}, \citenamefont {Yang},\ and\ \citenamefont
  {Zhu}}]{He:2025hin}%
  \BibitemOpen
  \bibfield  {author} {\bibinfo {author} {\bibfnamefont {C.-Q.}\ \bibnamefont
  {He}}, \bibinfo {author} {\bibfnamefont {H.}~\bibnamefont {Xing}}, \bibinfo
  {author} {\bibfnamefont {T.-Z.}\ \bibnamefont {Yang}}, \ and\ \bibinfo
  {author} {\bibfnamefont {H.~X.}\ \bibnamefont {Zhu}},\ }\href {\doibase
  10.1103/vtz2-6bnm} {\bibfield  {journal} {\bibinfo  {journal} {Phys. Rev.
  Lett.}\ }\textbf {\bibinfo {volume} {135}},\ \bibinfo {pages} {101901}
  (\bibinfo {year} {2025})},\ \Eprint {http://arxiv.org/abs/2503.20441}
  {arXiv:2503.20441 [hep-ph]} \BibitemShut {NoStop}%
\bibitem [{\citenamefont {Binnewies}\ \emph
  {et~al.}(1995{\natexlab{a}})\citenamefont {Binnewies}, \citenamefont
  {Kniehl},\ and\ \citenamefont {Kramer}}]{Binnewies:1994ju}%
  \BibitemOpen
  \bibfield  {author} {\bibinfo {author} {\bibfnamefont {J.}~\bibnamefont
  {Binnewies}}, \bibinfo {author} {\bibfnamefont {B.~A.}\ \bibnamefont
  {Kniehl}}, \ and\ \bibinfo {author} {\bibfnamefont {G.}~\bibnamefont
  {Kramer}},\ }\href {\doibase 10.1007/BF01556135} {\bibfield  {journal}
  {\bibinfo  {journal} {Z. Phys. C}\ }\textbf {\bibinfo {volume} {65}},\
  \bibinfo {pages} {471} (\bibinfo {year} {1995}{\natexlab{a}})},\ \Eprint
  {http://arxiv.org/abs/hep-ph/9407347} {arXiv:hep-ph/9407347} \BibitemShut
  {NoStop}%
\bibitem [{\citenamefont {Binnewies}\ \emph
  {et~al.}(1995{\natexlab{b}})\citenamefont {Binnewies}, \citenamefont
  {Kniehl},\ and\ \citenamefont {Kramer}}]{Binnewies:1995pt}%
  \BibitemOpen
  \bibfield  {author} {\bibinfo {author} {\bibfnamefont {J.}~\bibnamefont
  {Binnewies}}, \bibinfo {author} {\bibfnamefont {B.~A.}\ \bibnamefont
  {Kniehl}}, \ and\ \bibinfo {author} {\bibfnamefont {G.}~\bibnamefont
  {Kramer}},\ }\href {\doibase 10.1103/PhysRevD.52.4947} {\bibfield  {journal}
  {\bibinfo  {journal} {Phys. Rev. D}\ }\textbf {\bibinfo {volume} {52}},\
  \bibinfo {pages} {4947} (\bibinfo {year} {1995}{\natexlab{b}})},\ \Eprint
  {http://arxiv.org/abs/hep-ph/9503464} {arXiv:hep-ph/9503464} \BibitemShut
  {NoStop}%
\bibitem [{\citenamefont {Kniehl}\ \emph {et~al.}(2000)\citenamefont {Kniehl},
  \citenamefont {Kramer},\ and\ \citenamefont {Potter}}]{Kniehl:2000fe}%
  \BibitemOpen
  \bibfield  {author} {\bibinfo {author} {\bibfnamefont {B.~A.}\ \bibnamefont
  {Kniehl}}, \bibinfo {author} {\bibfnamefont {G.}~\bibnamefont {Kramer}}, \
  and\ \bibinfo {author} {\bibfnamefont {B.}~\bibnamefont {Potter}},\ }\href
  {\doibase 10.1016/S0550-3213(00)00303-5} {\bibfield  {journal} {\bibinfo
  {journal} {Nucl. Phys. B}\ }\textbf {\bibinfo {volume} {582}},\ \bibinfo
  {pages} {514} (\bibinfo {year} {2000})},\ \Eprint
  {http://arxiv.org/abs/hep-ph/0010289} {arXiv:hep-ph/0010289} \BibitemShut
  {NoStop}%
\bibitem [{\citenamefont {Bourhis}\ \emph {et~al.}(2001)\citenamefont
  {Bourhis}, \citenamefont {Fontannaz}, \citenamefont {Guillet},\ and\
  \citenamefont {Werlen}}]{Bourhis:2000gs}%
  \BibitemOpen
  \bibfield  {author} {\bibinfo {author} {\bibfnamefont {L.}~\bibnamefont
  {Bourhis}}, \bibinfo {author} {\bibfnamefont {M.}~\bibnamefont {Fontannaz}},
  \bibinfo {author} {\bibfnamefont {J.~P.}\ \bibnamefont {Guillet}}, \ and\
  \bibinfo {author} {\bibfnamefont {M.}~\bibnamefont {Werlen}},\ }\href
  {\doibase 10.1007/s100520100579} {\bibfield  {journal} {\bibinfo  {journal}
  {Eur. Phys. J. C}\ }\textbf {\bibinfo {volume} {19}},\ \bibinfo {pages} {89}
  (\bibinfo {year} {2001})},\ \Eprint {http://arxiv.org/abs/hep-ph/0009101}
  {arXiv:hep-ph/0009101} \BibitemShut {NoStop}%
\bibitem [{\citenamefont {Kretzer}(2000)}]{Kretzer:2000yf}%
  \BibitemOpen
  \bibfield  {author} {\bibinfo {author} {\bibfnamefont {S.}~\bibnamefont
  {Kretzer}},\ }\href {\doibase 10.1103/PhysRevD.62.054001} {\bibfield
  {journal} {\bibinfo  {journal} {Phys. Rev. D}\ }\textbf {\bibinfo {volume}
  {62}},\ \bibinfo {pages} {054001} (\bibinfo {year} {2000})},\ \Eprint
  {http://arxiv.org/abs/hep-ph/0003177} {arXiv:hep-ph/0003177} \BibitemShut
  {NoStop}%
\bibitem [{\citenamefont {Kretzer}\ \emph {et~al.}(2001)\citenamefont
  {Kretzer}, \citenamefont {Leader},\ and\ \citenamefont
  {Christova}}]{Kretzer:2001pz}%
  \BibitemOpen
  \bibfield  {author} {\bibinfo {author} {\bibfnamefont {S.}~\bibnamefont
  {Kretzer}}, \bibinfo {author} {\bibfnamefont {E.}~\bibnamefont {Leader}}, \
  and\ \bibinfo {author} {\bibfnamefont {E.}~\bibnamefont {Christova}},\ }\href
  {\doibase 10.1007/s100520100815} {\bibfield  {journal} {\bibinfo  {journal}
  {Eur. Phys. J. C}\ }\textbf {\bibinfo {volume} {22}},\ \bibinfo {pages} {269}
  (\bibinfo {year} {2001})},\ \Eprint {http://arxiv.org/abs/hep-ph/0108055}
  {arXiv:hep-ph/0108055} \BibitemShut {NoStop}%
\bibitem [{\citenamefont {de~Florian}\ \emph {et~al.}(2007)\citenamefont
  {de~Florian}, \citenamefont {Sassot},\ and\ \citenamefont
  {Stratmann}}]{deFlorian:2007ekg}%
  \BibitemOpen
  \bibfield  {author} {\bibinfo {author} {\bibfnamefont {D.}~\bibnamefont
  {de~Florian}}, \bibinfo {author} {\bibfnamefont {R.}~\bibnamefont {Sassot}},
  \ and\ \bibinfo {author} {\bibfnamefont {M.}~\bibnamefont {Stratmann}},\
  }\href {\doibase 10.1103/PhysRevD.76.074033} {\bibfield  {journal} {\bibinfo
  {journal} {Phys. Rev. D}\ }\textbf {\bibinfo {volume} {76}},\ \bibinfo
  {pages} {074033} (\bibinfo {year} {2007})},\ \Eprint
  {http://arxiv.org/abs/0707.1506} {arXiv:0707.1506 [hep-ph]} \BibitemShut
  {NoStop}%
\bibitem [{\citenamefont {de~Florian}\ \emph {et~al.}(2015)\citenamefont
  {de~Florian}, \citenamefont {Sassot}, \citenamefont {Epele}, \citenamefont
  {Hern{\'a}ndez-Pinto},\ and\ \citenamefont {Stratmann}}]{deFlorian:2014xna}%
  \BibitemOpen
  \bibfield  {author} {\bibinfo {author} {\bibfnamefont {D.}~\bibnamefont
  {de~Florian}}, \bibinfo {author} {\bibfnamefont {R.}~\bibnamefont {Sassot}},
  \bibinfo {author} {\bibfnamefont {M.}~\bibnamefont {Epele}}, \bibinfo
  {author} {\bibfnamefont {R.~J.}\ \bibnamefont {Hern{\'a}ndez-Pinto}}, \ and\
  \bibinfo {author} {\bibfnamefont {M.}~\bibnamefont {Stratmann}},\ }\href
  {\doibase 10.1103/PhysRevD.91.014035} {\bibfield  {journal} {\bibinfo
  {journal} {Phys. Rev. D}\ }\textbf {\bibinfo {volume} {91}},\ \bibinfo
  {pages} {014035} (\bibinfo {year} {2015})},\ \Eprint
  {http://arxiv.org/abs/1410.6027} {arXiv:1410.6027 [hep-ph]} \BibitemShut
  {NoStop}%
\bibitem [{\citenamefont {de~Florian}\ \emph {et~al.}(2017)\citenamefont
  {de~Florian}, \citenamefont {Epele}, \citenamefont {Hernandez-Pinto},
  \citenamefont {Sassot},\ and\ \citenamefont {Stratmann}}]{deFlorian:2017lwf}%
  \BibitemOpen
  \bibfield  {author} {\bibinfo {author} {\bibfnamefont {D.}~\bibnamefont
  {de~Florian}}, \bibinfo {author} {\bibfnamefont {M.}~\bibnamefont {Epele}},
  \bibinfo {author} {\bibfnamefont {R.~J.}\ \bibnamefont {Hernandez-Pinto}},
  \bibinfo {author} {\bibfnamefont {R.}~\bibnamefont {Sassot}}, \ and\ \bibinfo
  {author} {\bibfnamefont {M.}~\bibnamefont {Stratmann}},\ }\href {\doibase
  10.1103/PhysRevD.95.094019} {\bibfield  {journal} {\bibinfo  {journal} {Phys.
  Rev. D}\ }\textbf {\bibinfo {volume} {95}},\ \bibinfo {pages} {094019}
  (\bibinfo {year} {2017})},\ \Eprint {http://arxiv.org/abs/1702.06353}
  {arXiv:1702.06353 [hep-ph]} \BibitemShut {NoStop}%
\bibitem [{\citenamefont {Borsa}\ \emph
  {et~al.}(2022{\natexlab{a}})\citenamefont {Borsa}, \citenamefont
  {de~Florian}, \citenamefont {Sassot},\ and\ \citenamefont
  {Stratmann}}]{Borsa:2021ran}%
  \BibitemOpen
  \bibfield  {author} {\bibinfo {author} {\bibfnamefont {I.}~\bibnamefont
  {Borsa}}, \bibinfo {author} {\bibfnamefont {D.}~\bibnamefont {de~Florian}},
  \bibinfo {author} {\bibfnamefont {R.}~\bibnamefont {Sassot}}, \ and\ \bibinfo
  {author} {\bibfnamefont {M.}~\bibnamefont {Stratmann}},\ }\href {\doibase
  10.1103/PhysRevD.105.L031502} {\bibfield  {journal} {\bibinfo  {journal}
  {Phys. Rev. D}\ }\textbf {\bibinfo {volume} {105}},\ \bibinfo {pages}
  {L031502} (\bibinfo {year} {2022}{\natexlab{a}})},\ \Eprint
  {http://arxiv.org/abs/2110.14015} {arXiv:2110.14015 [hep-ph]} \BibitemShut
  {NoStop}%
\bibitem [{\citenamefont {Hirai}\ \emph {et~al.}(2007)\citenamefont {Hirai},
  \citenamefont {Kumano}, \citenamefont {Nagai},\ and\ \citenamefont
  {Sudoh}}]{Hirai:2007cx}%
  \BibitemOpen
  \bibfield  {author} {\bibinfo {author} {\bibfnamefont {M.}~\bibnamefont
  {Hirai}}, \bibinfo {author} {\bibfnamefont {S.}~\bibnamefont {Kumano}},
  \bibinfo {author} {\bibfnamefont {T.~H.}\ \bibnamefont {Nagai}}, \ and\
  \bibinfo {author} {\bibfnamefont {K.}~\bibnamefont {Sudoh}},\ }\href
  {\doibase 10.1103/PhysRevD.75.094009} {\bibfield  {journal} {\bibinfo
  {journal} {Phys. Rev. D}\ }\textbf {\bibinfo {volume} {75}},\ \bibinfo
  {pages} {094009} (\bibinfo {year} {2007})},\ \Eprint
  {http://arxiv.org/abs/hep-ph/0702250} {arXiv:hep-ph/0702250} \BibitemShut
  {NoStop}%
\bibitem [{\citenamefont {Albino}\ \emph {et~al.}(2008)\citenamefont {Albino},
  \citenamefont {Kniehl},\ and\ \citenamefont {Kramer}}]{Albino:2008fy}%
  \BibitemOpen
  \bibfield  {author} {\bibinfo {author} {\bibfnamefont {S.}~\bibnamefont
  {Albino}}, \bibinfo {author} {\bibfnamefont {B.~A.}\ \bibnamefont {Kniehl}},
  \ and\ \bibinfo {author} {\bibfnamefont {G.}~\bibnamefont {Kramer}},\ }\href
  {\doibase 10.1016/j.nuclphysb.2008.05.017} {\bibfield  {journal} {\bibinfo
  {journal} {Nucl. Phys. B}\ }\textbf {\bibinfo {volume} {803}},\ \bibinfo
  {pages} {42} (\bibinfo {year} {2008})},\ \Eprint
  {http://arxiv.org/abs/0803.2768} {arXiv:0803.2768 [hep-ph]} \BibitemShut
  {NoStop}%
\bibitem [{\citenamefont {Bertone}\ \emph {et~al.}(2018)\citenamefont
  {Bertone}, \citenamefont {Hartland}, \citenamefont {Nocera}, \citenamefont
  {Rojo},\ and\ \citenamefont {Rottoli}}]{Bertone:2018ecm}%
  \BibitemOpen
  \bibfield  {author} {\bibinfo {author} {\bibfnamefont {V.}~\bibnamefont
  {Bertone}}, \bibinfo {author} {\bibfnamefont {N.~P.}\ \bibnamefont
  {Hartland}}, \bibinfo {author} {\bibfnamefont {E.~R.}\ \bibnamefont
  {Nocera}}, \bibinfo {author} {\bibfnamefont {J.}~\bibnamefont {Rojo}}, \ and\
  \bibinfo {author} {\bibfnamefont {L.}~\bibnamefont {Rottoli}} (\bibinfo
  {collaboration} {NNPDF}),\ }\href {\doibase 10.1140/epjc/s10052-018-6130-4}
  {\bibfield  {journal} {\bibinfo  {journal} {Eur. Phys. J. C}\ }\textbf
  {\bibinfo {volume} {78}},\ \bibinfo {pages} {651} (\bibinfo {year} {2018})},\
  \bibinfo {note} {[Erratum: Eur.Phys.J.C 84, 155 (2024)]},\ \Eprint
  {http://arxiv.org/abs/1807.03310} {arXiv:1807.03310 [hep-ph]} \BibitemShut
  {NoStop}%
\bibitem [{\citenamefont {Khalek}\ \emph {et~al.}(2021)\citenamefont {Khalek},
  \citenamefont {Bertone},\ and\ \citenamefont {Nocera}}]{Khalek:2021gxf}%
  \BibitemOpen
  \bibfield  {author} {\bibinfo {author} {\bibfnamefont {R.~A.}\ \bibnamefont
  {Khalek}}, \bibinfo {author} {\bibfnamefont {V.}~\bibnamefont {Bertone}}, \
  and\ \bibinfo {author} {\bibfnamefont {E.~R.}\ \bibnamefont {Nocera}}
  (\bibinfo {collaboration} {MAP (Multi-dimensional Analyses of Partonic
  distributions)}),\ }\href {\doibase 10.1103/PhysRevD.104.034007} {\bibfield
  {journal} {\bibinfo  {journal} {Phys. Rev. D}\ }\textbf {\bibinfo {volume}
  {104}},\ \bibinfo {pages} {034007} (\bibinfo {year} {2021})},\ \Eprint
  {http://arxiv.org/abs/2105.08725} {arXiv:2105.08725 [hep-ph]} \BibitemShut
  {NoStop}%
\bibitem [{\citenamefont {Moffat}\ \emph {et~al.}(2021)\citenamefont {Moffat},
  \citenamefont {Melnitchouk}, \citenamefont {Rogers},\ and\ \citenamefont
  {Sato}}]{Moffat:2021dji}%
  \BibitemOpen
  \bibfield  {author} {\bibinfo {author} {\bibfnamefont {E.}~\bibnamefont
  {Moffat}}, \bibinfo {author} {\bibfnamefont {W.}~\bibnamefont {Melnitchouk}},
  \bibinfo {author} {\bibfnamefont {T.~C.}\ \bibnamefont {Rogers}}, \ and\
  \bibinfo {author} {\bibfnamefont {N.}~\bibnamefont {Sato}} (\bibinfo
  {collaboration} {Jefferson Lab Angular Momentum (JAM)}),\ }\href {\doibase
  10.1103/PhysRevD.104.016015} {\bibfield  {journal} {\bibinfo  {journal}
  {Phys. Rev. D}\ }\textbf {\bibinfo {volume} {104}},\ \bibinfo {pages}
  {016015} (\bibinfo {year} {2021})},\ \Eprint
  {http://arxiv.org/abs/2101.04664} {arXiv:2101.04664 [hep-ph]} \BibitemShut
  {NoStop}%
\bibitem [{\citenamefont {Gao}\ \emph {et~al.}(2024{\natexlab{a}})\citenamefont
  {Gao}, \citenamefont {Liu}, \citenamefont {Shen}, \citenamefont {Xing},\ and\
  \citenamefont {Zhao}}]{Gao:2024dbv}%
  \BibitemOpen
  \bibfield  {author} {\bibinfo {author} {\bibfnamefont {J.}~\bibnamefont
  {Gao}}, \bibinfo {author} {\bibfnamefont {C.}~\bibnamefont {Liu}}, \bibinfo
  {author} {\bibfnamefont {X.}~\bibnamefont {Shen}}, \bibinfo {author}
  {\bibfnamefont {H.}~\bibnamefont {Xing}}, \ and\ \bibinfo {author}
  {\bibfnamefont {Y.}~\bibnamefont {Zhao}},\ }\href {\doibase
  10.1103/PhysRevD.110.114019} {\bibfield  {journal} {\bibinfo  {journal}
  {Phys. Rev. D}\ }\textbf {\bibinfo {volume} {110}},\ \bibinfo {pages}
  {114019} (\bibinfo {year} {2024}{\natexlab{a}})},\ \Eprint
  {http://arxiv.org/abs/2407.04422} {arXiv:2407.04422 [hep-ph]} \BibitemShut
  {NoStop}%
\bibitem [{\citenamefont {Gao}\ \emph {et~al.}(2024{\natexlab{b}})\citenamefont
  {Gao}, \citenamefont {Liu}, \citenamefont {Shen}, \citenamefont {Xing},\ and\
  \citenamefont {Zhao}}]{Gao:2024nkz}%
  \BibitemOpen
  \bibfield  {author} {\bibinfo {author} {\bibfnamefont {J.}~\bibnamefont
  {Gao}}, \bibinfo {author} {\bibfnamefont {C.}~\bibnamefont {Liu}}, \bibinfo
  {author} {\bibfnamefont {X.}~\bibnamefont {Shen}}, \bibinfo {author}
  {\bibfnamefont {H.}~\bibnamefont {Xing}}, \ and\ \bibinfo {author}
  {\bibfnamefont {Y.}~\bibnamefont {Zhao}},\ }\href {\doibase
  10.1103/PhysRevLett.132.261903} {\bibfield  {journal} {\bibinfo  {journal}
  {Phys. Rev. Lett.}\ }\textbf {\bibinfo {volume} {132}},\ \bibinfo {pages}
  {261903} (\bibinfo {year} {2024}{\natexlab{b}})},\ \Eprint
  {http://arxiv.org/abs/2401.02781} {arXiv:2401.02781 [hep-ph]} \BibitemShut
  {NoStop}%
\bibitem [{\citenamefont {Gao}\ \emph {et~al.}(2025{\natexlab{a}})\citenamefont
  {Gao}, \citenamefont {Liu}, \citenamefont {Li}, \citenamefont {Shen},
  \citenamefont {Xing}, \citenamefont {Zhao},\ and\ \citenamefont
  {Zhou}}]{Gao:2025bko}%
  \BibitemOpen
  \bibfield  {author} {\bibinfo {author} {\bibfnamefont {J.}~\bibnamefont
  {Gao}}, \bibinfo {author} {\bibfnamefont {C.}~\bibnamefont {Liu}}, \bibinfo
  {author} {\bibfnamefont {M.}~\bibnamefont {Li}}, \bibinfo {author}
  {\bibfnamefont {X.}~\bibnamefont {Shen}}, \bibinfo {author} {\bibfnamefont
  {H.}~\bibnamefont {Xing}}, \bibinfo {author} {\bibfnamefont {Y.}~\bibnamefont
  {Zhao}}, \ and\ \bibinfo {author} {\bibfnamefont {Y.}~\bibnamefont {Zhou}},\
  }\href {\doibase 10.1103/t5ds-vvc4} {\bibfield  {journal} {\bibinfo
  {journal} {Phys. Rev. D}\ }\textbf {\bibinfo {volume} {112}},\ \bibinfo
  {pages} {054045} (\bibinfo {year} {2025}{\natexlab{a}})},\ \Eprint
  {http://arxiv.org/abs/2503.21311} {arXiv:2503.21311 [hep-ph]} \BibitemShut
  {NoStop}%
\bibitem [{\citenamefont {Bertone}\ \emph {et~al.}(2017)\citenamefont
  {Bertone}, \citenamefont {Carrazza}, \citenamefont {Hartland}, \citenamefont
  {Nocera},\ and\ \citenamefont {Rojo}}]{Bertone:2017tyb}%
  \BibitemOpen
  \bibfield  {author} {\bibinfo {author} {\bibfnamefont {V.}~\bibnamefont
  {Bertone}}, \bibinfo {author} {\bibfnamefont {S.}~\bibnamefont {Carrazza}},
  \bibinfo {author} {\bibfnamefont {N.~P.}\ \bibnamefont {Hartland}}, \bibinfo
  {author} {\bibfnamefont {E.~R.}\ \bibnamefont {Nocera}}, \ and\ \bibinfo
  {author} {\bibfnamefont {J.}~\bibnamefont {Rojo}} (\bibinfo {collaboration}
  {NNPDF}),\ }\href {\doibase 10.1140/epjc/s10052-017-5088-y} {\bibfield
  {journal} {\bibinfo  {journal} {Eur. Phys. J. C}\ }\textbf {\bibinfo {volume}
  {77}},\ \bibinfo {pages} {516} (\bibinfo {year} {2017})},\ \Eprint
  {http://arxiv.org/abs/1706.07049} {arXiv:1706.07049 [hep-ph]} \BibitemShut
  {NoStop}%
\bibitem [{\citenamefont {Anderle}\ \emph {et~al.}(2015)\citenamefont
  {Anderle}, \citenamefont {Ringer},\ and\ \citenamefont
  {Stratmann}}]{Anderle:2015lqa}%
  \BibitemOpen
  \bibfield  {author} {\bibinfo {author} {\bibfnamefont {D.~P.}\ \bibnamefont
  {Anderle}}, \bibinfo {author} {\bibfnamefont {F.}~\bibnamefont {Ringer}}, \
  and\ \bibinfo {author} {\bibfnamefont {M.}~\bibnamefont {Stratmann}},\ }\href
  {\doibase 10.1103/PhysRevD.92.114017} {\bibfield  {journal} {\bibinfo
  {journal} {Phys. Rev. D}\ }\textbf {\bibinfo {volume} {92}},\ \bibinfo
  {pages} {114017} (\bibinfo {year} {2015})},\ \Eprint
  {http://arxiv.org/abs/1510.05845} {arXiv:1510.05845 [hep-ph]} \BibitemShut
  {NoStop}%
\bibitem [{\citenamefont {Abdolmaleki}\ \emph {et~al.}(2021)\citenamefont
  {Abdolmaleki}, \citenamefont {Soleymaninia}, \citenamefont {Khanpour},
  \citenamefont {Amoroso}, \citenamefont {Giuli}, \citenamefont {Glazov},
  \citenamefont {Luszczak}, \citenamefont {Olness},\ and\ \citenamefont
  {Zenaiev}}]{Abdolmaleki:2021yjf}%
  \BibitemOpen
  \bibfield  {author} {\bibinfo {author} {\bibfnamefont {H.}~\bibnamefont
  {Abdolmaleki}}, \bibinfo {author} {\bibfnamefont {M.}~\bibnamefont
  {Soleymaninia}}, \bibinfo {author} {\bibfnamefont {H.}~\bibnamefont
  {Khanpour}}, \bibinfo {author} {\bibfnamefont {S.}~\bibnamefont {Amoroso}},
  \bibinfo {author} {\bibfnamefont {F.}~\bibnamefont {Giuli}}, \bibinfo
  {author} {\bibfnamefont {A.}~\bibnamefont {Glazov}}, \bibinfo {author}
  {\bibfnamefont {A.}~\bibnamefont {Luszczak}}, \bibinfo {author}
  {\bibfnamefont {F.}~\bibnamefont {Olness}}, \ and\ \bibinfo {author}
  {\bibfnamefont {O.}~\bibnamefont {Zenaiev}} (\bibinfo {collaboration}
  {xfitter Developers{\textquoteright} Team}),\ }\href {\doibase
  10.1103/PhysRevD.104.056019} {\bibfield  {journal} {\bibinfo  {journal}
  {Phys. Rev. D}\ }\textbf {\bibinfo {volume} {104}},\ \bibinfo {pages}
  {056019} (\bibinfo {year} {2021})},\ \Eprint
  {http://arxiv.org/abs/2105.11306} {arXiv:2105.11306 [hep-ph]} \BibitemShut
  {NoStop}%
\bibitem [{\citenamefont {Borsa}\ \emph
  {et~al.}(2022{\natexlab{b}})\citenamefont {Borsa}, \citenamefont {Sassot},
  \citenamefont {de~Florian}, \citenamefont {Stratmann},\ and\ \citenamefont
  {Vogelsang}}]{Borsa:2022vvp}%
  \BibitemOpen
  \bibfield  {author} {\bibinfo {author} {\bibfnamefont {I.}~\bibnamefont
  {Borsa}}, \bibinfo {author} {\bibfnamefont {R.}~\bibnamefont {Sassot}},
  \bibinfo {author} {\bibfnamefont {D.}~\bibnamefont {de~Florian}}, \bibinfo
  {author} {\bibfnamefont {M.}~\bibnamefont {Stratmann}}, \ and\ \bibinfo
  {author} {\bibfnamefont {W.}~\bibnamefont {Vogelsang}},\ }\href {\doibase
  10.1103/PhysRevLett.129.012002} {\bibfield  {journal} {\bibinfo  {journal}
  {Phys. Rev. Lett.}\ }\textbf {\bibinfo {volume} {129}},\ \bibinfo {pages}
  {012002} (\bibinfo {year} {2022}{\natexlab{b}})},\ \Eprint
  {http://arxiv.org/abs/2202.05060} {arXiv:2202.05060 [hep-ph]} \BibitemShut
  {NoStop}%
\bibitem [{\citenamefont {Gao}\ \emph {et~al.}(2025{\natexlab{b}})\citenamefont
  {Gao}, \citenamefont {Shen}, \citenamefont {Xing}, \citenamefont {Zhao},\
  and\ \citenamefont {Zhou}}]{Gao:2025hlm}%
  \BibitemOpen
  \bibfield  {author} {\bibinfo {author} {\bibfnamefont {J.}~\bibnamefont
  {Gao}}, \bibinfo {author} {\bibfnamefont {X.}~\bibnamefont {Shen}}, \bibinfo
  {author} {\bibfnamefont {H.}~\bibnamefont {Xing}}, \bibinfo {author}
  {\bibfnamefont {Y.}~\bibnamefont {Zhao}}, \ and\ \bibinfo {author}
  {\bibfnamefont {B.}~\bibnamefont {Zhou}},\ }\href {\doibase
  10.1103/mcwy-b221} {\bibfield  {journal} {\bibinfo  {journal} {Phys. Rev.
  Lett.}\ }\textbf {\bibinfo {volume} {135}},\ \bibinfo {pages} {041902}
  (\bibinfo {year} {2025}{\natexlab{b}})},\ \Eprint
  {http://arxiv.org/abs/2502.17837} {arXiv:2502.17837 [hep-ph]} \BibitemShut
  {NoStop}%
\bibitem [{\citenamefont {Abdul~Khalek}\ \emph {et~al.}(2022)\citenamefont
  {Abdul~Khalek}, \citenamefont {Bertone}, \citenamefont {Khoudli},\ and\
  \citenamefont {Nocera}}]{AbdulKhalek:2022laj}%
  \BibitemOpen
  \bibfield  {author} {\bibinfo {author} {\bibfnamefont {R.}~\bibnamefont
  {Abdul~Khalek}}, \bibinfo {author} {\bibfnamefont {V.}~\bibnamefont
  {Bertone}}, \bibinfo {author} {\bibfnamefont {A.}~\bibnamefont {Khoudli}}, \
  and\ \bibinfo {author} {\bibfnamefont {E.~R.}\ \bibnamefont {Nocera}},\
  }\href {\doibase 10.1016/j.physletb.2022.137456} {\bibfield  {journal}
  {\bibinfo  {journal} {Phys. Lett. B}\ }\textbf {\bibinfo {volume} {834}},\
  \bibinfo {pages} {137456} (\bibinfo {year} {2022})},\ \Eprint
  {http://arxiv.org/abs/2204.10331} {arXiv:2204.10331 [hep-ph]} \BibitemShut
  {NoStop}%
\bibitem [{\citenamefont {Li}\ \emph {et~al.}(2025)\citenamefont {Li},
  \citenamefont {Anderle}, \citenamefont {Xing},\ and\ \citenamefont
  {Zhao}}]{Li:2024etc}%
  \BibitemOpen
  \bibfield  {author} {\bibinfo {author} {\bibfnamefont {M.}~\bibnamefont
  {Li}}, \bibinfo {author} {\bibfnamefont {D.~P.}\ \bibnamefont {Anderle}},
  \bibinfo {author} {\bibfnamefont {H.}~\bibnamefont {Xing}}, \ and\ \bibinfo
  {author} {\bibfnamefont {Y.}~\bibnamefont {Zhao}},\ }\href {\doibase
  10.1103/PhysRevD.111.034030} {\bibfield  {journal} {\bibinfo  {journal}
  {Phys. Rev. D}\ }\textbf {\bibinfo {volume} {111}},\ \bibinfo {pages}
  {034030} (\bibinfo {year} {2025})},\ \Eprint
  {http://arxiv.org/abs/2404.11527} {arXiv:2404.11527 [hep-ph]} \BibitemShut
  {NoStop}%
\bibitem [{\citenamefont {de~Florian}\ \emph {et~al.}(1998)\citenamefont
  {de~Florian}, \citenamefont {Stratmann},\ and\ \citenamefont
  {Vogelsang}}]{deFlorian:1997zj}%
  \BibitemOpen
  \bibfield  {author} {\bibinfo {author} {\bibfnamefont {D.}~\bibnamefont
  {de~Florian}}, \bibinfo {author} {\bibfnamefont {M.}~\bibnamefont
  {Stratmann}}, \ and\ \bibinfo {author} {\bibfnamefont {W.}~\bibnamefont
  {Vogelsang}},\ }\href {\doibase 10.1103/PhysRevD.57.5811} {\bibfield
  {journal} {\bibinfo  {journal} {Phys. Rev. D}\ }\textbf {\bibinfo {volume}
  {57}},\ \bibinfo {pages} {5811} (\bibinfo {year} {1998})},\ \Eprint
  {http://arxiv.org/abs/hep-ph/9711387} {arXiv:hep-ph/9711387} \BibitemShut
  {NoStop}%
\bibitem [{\citenamefont {Bonino}\ \emph
  {et~al.}(2024{\natexlab{a}})\citenamefont {Bonino}, \citenamefont
  {Gehrmann},\ and\ \citenamefont {Stagnitto}}]{Bonino:2024qbh}%
  \BibitemOpen
  \bibfield  {author} {\bibinfo {author} {\bibfnamefont {L.}~\bibnamefont
  {Bonino}}, \bibinfo {author} {\bibfnamefont {T.}~\bibnamefont {Gehrmann}}, \
  and\ \bibinfo {author} {\bibfnamefont {G.}~\bibnamefont {Stagnitto}},\ }\href
  {\doibase 10.1103/PhysRevLett.132.251901} {\bibfield  {journal} {\bibinfo
  {journal} {Phys. Rev. Lett.}\ }\textbf {\bibinfo {volume} {132}},\ \bibinfo
  {pages} {251901} (\bibinfo {year} {2024}{\natexlab{a}})},\ \Eprint
  {http://arxiv.org/abs/2401.16281} {arXiv:2401.16281 [hep-ph]} \BibitemShut
  {NoStop}%
\bibitem [{\citenamefont {Goyal}\ \emph {et~al.}(2024)\citenamefont {Goyal},
  \citenamefont {Moch}, \citenamefont {Pathak}, \citenamefont {Rana},\ and\
  \citenamefont {Ravindran}}]{Goyal:2023zdi}%
  \BibitemOpen
  \bibfield  {author} {\bibinfo {author} {\bibfnamefont {S.}~\bibnamefont
  {Goyal}}, \bibinfo {author} {\bibfnamefont {S.-O.}\ \bibnamefont {Moch}},
  \bibinfo {author} {\bibfnamefont {V.}~\bibnamefont {Pathak}}, \bibinfo
  {author} {\bibfnamefont {N.}~\bibnamefont {Rana}}, \ and\ \bibinfo {author}
  {\bibfnamefont {V.}~\bibnamefont {Ravindran}},\ }\href {\doibase
  10.1103/PhysRevLett.132.251902} {\bibfield  {journal} {\bibinfo  {journal}
  {Phys. Rev. Lett.}\ }\textbf {\bibinfo {volume} {132}},\ \bibinfo {pages}
  {251902} (\bibinfo {year} {2024})},\ \Eprint
  {http://arxiv.org/abs/2312.17711} {arXiv:2312.17711 [hep-ph]} \BibitemShut
  {NoStop}%
\bibitem [{\citenamefont {Aversa}\ \emph {et~al.}(1989)\citenamefont {Aversa},
  \citenamefont {Chiappetta}, \citenamefont {Greco},\ and\ \citenamefont
  {Guillet}}]{Aversa:1988vb}%
  \BibitemOpen
  \bibfield  {author} {\bibinfo {author} {\bibfnamefont {F.}~\bibnamefont
  {Aversa}}, \bibinfo {author} {\bibfnamefont {P.}~\bibnamefont {Chiappetta}},
  \bibinfo {author} {\bibfnamefont {M.}~\bibnamefont {Greco}}, \ and\ \bibinfo
  {author} {\bibfnamefont {J.~P.}\ \bibnamefont {Guillet}},\ }\href {\doibase
  10.1016/0550-3213(89)90288-5} {\bibfield  {journal} {\bibinfo  {journal}
  {Nucl. Phys. B}\ }\textbf {\bibinfo {volume} {327}},\ \bibinfo {pages} {105}
  (\bibinfo {year} {1989})}\BibitemShut {NoStop}%
\bibitem [{\citenamefont {Jäger}\ \emph {et~al.}(2003)\citenamefont {Jäger},
  \citenamefont {Schafer}, \citenamefont {Stratmann},\ and\ \citenamefont
  {Vogelsang}}]{Jager:2002xm}%
  \BibitemOpen
  \bibfield  {author} {\bibinfo {author} {\bibfnamefont {B.}~\bibnamefont
  {Jäger}}, \bibinfo {author} {\bibfnamefont {A.}~\bibnamefont {Schafer}},
  \bibinfo {author} {\bibfnamefont {M.}~\bibnamefont {Stratmann}}, \ and\
  \bibinfo {author} {\bibfnamefont {W.}~\bibnamefont {Vogelsang}},\ }\href
  {\doibase 10.1103/PhysRevD.67.054005} {\bibfield  {journal} {\bibinfo
  {journal} {Phys. Rev. D}\ }\textbf {\bibinfo {volume} {67}},\ \bibinfo
  {pages} {054005} (\bibinfo {year} {2003})},\ \Eprint
  {http://arxiv.org/abs/hep-ph/0211007} {arXiv:hep-ph/0211007} \BibitemShut
  {NoStop}%
\bibitem [{\citenamefont {Czakon}\ \emph {et~al.}(2025)\citenamefont {Czakon},
  \citenamefont {Generet}, \citenamefont {Mitov},\ and\ \citenamefont
  {Poncelet}}]{Czakon:2025yti}%
  \BibitemOpen
  \bibfield  {author} {\bibinfo {author} {\bibfnamefont {M.}~\bibnamefont
  {Czakon}}, \bibinfo {author} {\bibfnamefont {T.}~\bibnamefont {Generet}},
  \bibinfo {author} {\bibfnamefont {A.}~\bibnamefont {Mitov}}, \ and\ \bibinfo
  {author} {\bibfnamefont {R.}~\bibnamefont {Poncelet}},\ }\href {\doibase
  10.1103/mhkl-vt96} {\bibfield  {journal} {\bibinfo  {journal} {Phys. Rev.
  Lett.}\ }\textbf {\bibinfo {volume} {135}},\ \bibinfo {pages} {17} (\bibinfo
  {year} {2025})},\ \Eprint {http://arxiv.org/abs/2503.11489} {arXiv:2503.11489
  [hep-ph]} \BibitemShut {NoStop}%
\bibitem [{\citenamefont {Generet}\ \emph {et~al.}(2026)\citenamefont
  {Generet}, \citenamefont {Poncelet},\ and\ \citenamefont
  {Mu{\v{s}}kinja}}]{Generet:2025bqx}%
  \BibitemOpen
  \bibfield  {author} {\bibinfo {author} {\bibfnamefont {T.}~\bibnamefont
  {Generet}}, \bibinfo {author} {\bibfnamefont {R.}~\bibnamefont {Poncelet}}, \
  and\ \bibinfo {author} {\bibfnamefont {M.}~\bibnamefont {Mu{\v{s}}kinja}},\
  }\href {\doibase 10.1007/JHEP02(2026)023} {\bibfield  {journal} {\bibinfo
  {journal} {JHEP}\ }\textbf {\bibinfo {volume} {02}},\ \bibinfo {pages} {023}
  (\bibinfo {year} {2026})},\ \Eprint {http://arxiv.org/abs/2510.24525}
  {arXiv:2510.24525 [hep-ph]} \BibitemShut {NoStop}%
\bibitem [{\citenamefont {Benedikt}\ \emph {et~al.}(2025)\citenamefont
  {Benedikt} \emph {et~al.}}]{FCC:2025lpp}%
  \BibitemOpen
  \bibfield  {author} {\bibinfo {author} {\bibfnamefont {M.}~\bibnamefont
  {Benedikt}} \emph {et~al.} (\bibinfo {collaboration} {FCC}),\ }\href
  {\doibase 10.1140/epjc/s10052-025-15077-x} {\bibfield  {journal} {\bibinfo
  {journal} {Eur. Phys. J. C}\ }\textbf {\bibinfo {volume} {85}},\ \bibinfo
  {pages} {1468} (\bibinfo {year} {2025})},\ \Eprint
  {http://arxiv.org/abs/2505.00272} {arXiv:2505.00272 [hep-ex]} \BibitemShut
  {NoStop}%
\bibitem [{\citenamefont {Zhou}\ and\ \citenamefont
  {Gao}(2025)}]{Zhou:2024cyk}%
  \BibitemOpen
  \bibfield  {author} {\bibinfo {author} {\bibfnamefont {B.}~\bibnamefont
  {Zhou}}\ and\ \bibinfo {author} {\bibfnamefont {J.}~\bibnamefont {Gao}},\
  }\href {\doibase 10.1007/JHEP02(2025)003} {\bibfield  {journal} {\bibinfo
  {journal} {JHEP}\ }\textbf {\bibinfo {volume} {02}},\ \bibinfo {pages} {003}
  (\bibinfo {year} {2025})},\ \Eprint {http://arxiv.org/abs/2407.10059}
  {arXiv:2407.10059 [hep-ph]} \BibitemShut {NoStop}%
\bibitem [{\citenamefont {d'Enterria}\ \emph {et~al.}(2025)\citenamefont
  {d'Enterria}, \citenamefont {Monni}, \citenamefont {Skands},\ and\
  \citenamefont {Verbytskyi}}]{dEnterria:2025pml}%
  \BibitemOpen
  \bibfield  {author} {\bibinfo {author} {\bibfnamefont {D.}~\bibnamefont
  {d'Enterria}}, \bibinfo {author} {\bibfnamefont {P.~F.}\ \bibnamefont
  {Monni}}, \bibinfo {author} {\bibfnamefont {P.}~\bibnamefont {Skands}}, \
  and\ \bibinfo {author} {\bibfnamefont {A.}~\bibnamefont {Verbytskyi}}\
  }(\bibinfo {year} {2025})\ \Eprint {http://arxiv.org/abs/2503.23855}
  {arXiv:2503.23855 [hep-ex]} \BibitemShut {NoStop}%
\bibitem [{\citenamefont {Gehrmann-De~Ridder}\ \emph
  {et~al.}(2005)\citenamefont {Gehrmann-De~Ridder}, \citenamefont {Gehrmann},\
  and\ \citenamefont {Glover}}]{Gehrmann-DeRidder:2005btv}%
  \BibitemOpen
  \bibfield  {author} {\bibinfo {author} {\bibfnamefont {A.}~\bibnamefont
  {Gehrmann-De~Ridder}}, \bibinfo {author} {\bibfnamefont {T.}~\bibnamefont
  {Gehrmann}}, \ and\ \bibinfo {author} {\bibfnamefont {E.~W.~N.}\ \bibnamefont
  {Glover}},\ }\href {\doibase 10.1088/1126-6708/2005/09/056} {\bibfield
  {journal} {\bibinfo  {journal} {JHEP}\ }\textbf {\bibinfo {volume} {09}},\
  \bibinfo {pages} {056} (\bibinfo {year} {2005})},\ \Eprint
  {http://arxiv.org/abs/hep-ph/0505111} {arXiv:hep-ph/0505111} \BibitemShut
  {NoStop}%
\bibitem [{\citenamefont {Currie}\ \emph {et~al.}(2013)\citenamefont {Currie},
  \citenamefont {Glover},\ and\ \citenamefont {Wells}}]{Currie:2013vh}%
  \BibitemOpen
  \bibfield  {author} {\bibinfo {author} {\bibfnamefont {J.}~\bibnamefont
  {Currie}}, \bibinfo {author} {\bibfnamefont {E.}~\bibnamefont {Glover}}, \
  and\ \bibinfo {author} {\bibfnamefont {S.}~\bibnamefont {Wells}},\ }\href
  {\doibase 10.1007/JHEP04(2013)066} {\bibfield  {journal} {\bibinfo  {journal}
  {JHEP}\ }\textbf {\bibinfo {volume} {04}},\ \bibinfo {pages} {066} (\bibinfo
  {year} {2013})},\ \Eprint {http://arxiv.org/abs/1301.4693} {arXiv:1301.4693
  [hep-ph]} \BibitemShut {NoStop}%
\bibitem [{\citenamefont {Gehrmann}\ and\ \citenamefont
  {Sch\"urmann}(2022)}]{Gehrmann:2022cih}%
  \BibitemOpen
  \bibfield  {author} {\bibinfo {author} {\bibfnamefont {T.}~\bibnamefont
  {Gehrmann}}\ and\ \bibinfo {author} {\bibfnamefont {R.}~\bibnamefont
  {Sch\"urmann}},\ }\href {\doibase 10.1007/JHEP04(2022)031} {\bibfield
  {journal} {\bibinfo  {journal} {JHEP}\ }\textbf {\bibinfo {volume} {04}},\
  \bibinfo {pages} {031} (\bibinfo {year} {2022})},\ \Eprint
  {http://arxiv.org/abs/2201.06982} {arXiv:2201.06982 [hep-ph]} \BibitemShut
  {NoStop}%
\bibitem [{\citenamefont {Gehrmann}\ and\ \citenamefont
  {Stagnitto}(2022)}]{Gehrmann:2022pzd}%
  \BibitemOpen
  \bibfield  {author} {\bibinfo {author} {\bibfnamefont {T.}~\bibnamefont
  {Gehrmann}}\ and\ \bibinfo {author} {\bibfnamefont {G.}~\bibnamefont
  {Stagnitto}},\ }\href {\doibase 10.1007/JHEP10(2022)136} {\bibfield
  {journal} {\bibinfo  {journal} {JHEP}\ }\textbf {\bibinfo {volume} {10}},\
  \bibinfo {pages} {136} (\bibinfo {year} {2022})},\ \Eprint
  {http://arxiv.org/abs/2208.02650} {arXiv:2208.02650 [hep-ph]} \BibitemShut
  {NoStop}%
\bibitem [{\citenamefont {Bonino}\ \emph
  {et~al.}(2024{\natexlab{b}})\citenamefont {Bonino}, \citenamefont {Gehrmann},
  \citenamefont {Marcoli}, \citenamefont {Sch\"urmann},\ and\ \citenamefont
  {Stagnitto}}]{Bonino:2024adk}%
  \BibitemOpen
  \bibfield  {author} {\bibinfo {author} {\bibfnamefont {L.}~\bibnamefont
  {Bonino}}, \bibinfo {author} {\bibfnamefont {T.}~\bibnamefont {Gehrmann}},
  \bibinfo {author} {\bibfnamefont {M.}~\bibnamefont {Marcoli}}, \bibinfo
  {author} {\bibfnamefont {R.}~\bibnamefont {Sch\"urmann}}, \ and\ \bibinfo
  {author} {\bibfnamefont {G.}~\bibnamefont {Stagnitto}},\ }\href {\doibase
  10.1007/JHEP08(2024)073} {\bibfield  {journal} {\bibinfo  {journal} {JHEP}\
  }\textbf {\bibinfo {volume} {08}},\ \bibinfo {pages} {073} (\bibinfo {year}
  {2024}{\natexlab{b}})},\ \Eprint {http://arxiv.org/abs/2406.09925}
  {arXiv:2406.09925 [hep-ph]} \BibitemShut {NoStop}%
\bibitem [{\citenamefont {Caletti}\ \emph {et~al.}(2024)\citenamefont
  {Caletti}, \citenamefont {Gehrmann-De~Ridder}, \citenamefont {Huss},
  \citenamefont {Garcia},\ and\ \citenamefont {Stagnitto}}]{Caletti:2024xaw}%
  \BibitemOpen
  \bibfield  {author} {\bibinfo {author} {\bibfnamefont {S.}~\bibnamefont
  {Caletti}}, \bibinfo {author} {\bibfnamefont {A.}~\bibnamefont
  {Gehrmann-De~Ridder}}, \bibinfo {author} {\bibfnamefont {A.}~\bibnamefont
  {Huss}}, \bibinfo {author} {\bibfnamefont {A.~R.}\ \bibnamefont {Garcia}}, \
  and\ \bibinfo {author} {\bibfnamefont {G.}~\bibnamefont {Stagnitto}},\ }\href
  {\doibase 10.1007/JHEP10(2024)027} {\bibfield  {journal} {\bibinfo  {journal}
  {JHEP}\ }\textbf {\bibinfo {volume} {10}},\ \bibinfo {pages} {027} (\bibinfo
  {year} {2024})},\ \Eprint {http://arxiv.org/abs/2405.17540} {arXiv:2405.17540
  [hep-ph]} \BibitemShut {NoStop}%
\bibitem [{\citenamefont {Huss}\ \emph {et~al.}(2025)\citenamefont {Huss} \emph
  {et~al.}}]{NNLOJET:2025rno}%
  \BibitemOpen
  \bibfield  {author} {\bibinfo {author} {\bibfnamefont {A.}~\bibnamefont
  {Huss}} \emph {et~al.} (\bibinfo {collaboration} {NNLOJET}),\ }\href@noop {}
  {\  (\bibinfo {year} {2025})},\ \Eprint {http://arxiv.org/abs/2503.22804}
  {arXiv:2503.22804 [hep-ph]} \BibitemShut {NoStop}%
\bibitem [{\citenamefont {Currie}\ \emph {et~al.}(2017)\citenamefont {Currie},
  \citenamefont {Gehrmann}, \citenamefont {Huss},\ and\ \citenamefont
  {Niehues}}]{Currie:2017tpe}%
  \BibitemOpen
  \bibfield  {author} {\bibinfo {author} {\bibfnamefont {J.}~\bibnamefont
  {Currie}}, \bibinfo {author} {\bibfnamefont {T.}~\bibnamefont {Gehrmann}},
  \bibinfo {author} {\bibfnamefont {A.}~\bibnamefont {Huss}}, \ and\ \bibinfo
  {author} {\bibfnamefont {J.}~\bibnamefont {Niehues}},\ }\href {\doibase
  10.1007/JHEP07(2017)018} {\bibfield  {journal} {\bibinfo  {journal} {JHEP}\
  }\textbf {\bibinfo {volume} {07}},\ \bibinfo {pages} {018} (\bibinfo {year}
  {2017})},\ \bibinfo {note} {[Erratum: JHEP 12, 042 (2020)]},\ \Eprint
  {http://arxiv.org/abs/1703.05977} {arXiv:1703.05977 [hep-ph]} \BibitemShut
  {NoStop}%
\bibitem [{\citenamefont {Bonino}\ \emph
  {et~al.}(2024{\natexlab{c}})\citenamefont {Bonino}, \citenamefont
  {Cacciari},\ and\ \citenamefont {Stagnitto}}]{Bonino:2023icn}%
  \BibitemOpen
  \bibfield  {author} {\bibinfo {author} {\bibfnamefont {L.}~\bibnamefont
  {Bonino}}, \bibinfo {author} {\bibfnamefont {M.}~\bibnamefont {Cacciari}}, \
  and\ \bibinfo {author} {\bibfnamefont {G.}~\bibnamefont {Stagnitto}},\ }\href
  {\doibase 10.1007/JHEP06(2024)040} {\bibfield  {journal} {\bibinfo  {journal}
  {JHEP}\ }\textbf {\bibinfo {volume} {06}},\ \bibinfo {pages} {040} (\bibinfo
  {year} {2024}{\natexlab{c}})},\ \Eprint {http://arxiv.org/abs/2312.12519}
  {arXiv:2312.12519 [hep-ph]} \BibitemShut {NoStop}%
\bibitem [{\citenamefont {Barate}\ \emph {et~al.}(2000)\citenamefont {Barate}
  \emph {et~al.}}]{ALEPH:1999udi}%
  \BibitemOpen
  \bibfield  {author} {\bibinfo {author} {\bibfnamefont {R.}~\bibnamefont
  {Barate}} \emph {et~al.} (\bibinfo {collaboration} {ALEPH}),\ }\href
  {\doibase 10.1007/s100520000443} {\bibfield  {journal} {\bibinfo  {journal}
  {Eur. Phys. J. C}\ }\textbf {\bibinfo {volume} {16}},\ \bibinfo {pages} {613}
  (\bibinfo {year} {2000})}\BibitemShut {NoStop}%
\bibitem [{\citenamefont {Acciarri}\ \emph {et~al.}(1996)\citenamefont
  {Acciarri} \emph {et~al.}}]{L3:1995rnv}%
  \BibitemOpen
  \bibfield  {author} {\bibinfo {author} {\bibfnamefont {M.}~\bibnamefont
  {Acciarri}} \emph {et~al.} (\bibinfo {collaboration} {L3}),\ }\href {\doibase
  10.1016/0370-2693(96)00095-0} {\bibfield  {journal} {\bibinfo  {journal}
  {Phys. Lett. B}\ }\textbf {\bibinfo {volume} {371}},\ \bibinfo {pages} {126}
  (\bibinfo {year} {1996})}\BibitemShut {NoStop}%
\bibitem [{\citenamefont {Abbiendi}\ \emph {et~al.}(2004)\citenamefont
  {Abbiendi} \emph {et~al.}}]{OPAL:2004prv}%
  \BibitemOpen
  \bibfield  {author} {\bibinfo {author} {\bibfnamefont {G.}~\bibnamefont
  {Abbiendi}} \emph {et~al.} (\bibinfo {collaboration} {OPAL}),\ }\href
  {\doibase 10.1140/epjc/s2004-01964-4} {\bibfield  {journal} {\bibinfo
  {journal} {Eur. Phys. J. C}\ }\textbf {\bibinfo {volume} {37}},\ \bibinfo
  {pages} {25} (\bibinfo {year} {2004})},\ \Eprint
  {http://arxiv.org/abs/hep-ex/0404026} {arXiv:hep-ex/0404026} \BibitemShut
  {NoStop}%
\bibitem [{\citenamefont {Catani}\ \emph {et~al.}(1991)\citenamefont {Catani},
  \citenamefont {Dokshitzer}, \citenamefont {Olsson}, \citenamefont {Turnock},\
  and\ \citenamefont {Webber}}]{Catani:1991hj}%
  \BibitemOpen
  \bibfield  {author} {\bibinfo {author} {\bibfnamefont {S.}~\bibnamefont
  {Catani}}, \bibinfo {author} {\bibfnamefont {Y.~L.}\ \bibnamefont
  {Dokshitzer}}, \bibinfo {author} {\bibfnamefont {M.}~\bibnamefont {Olsson}},
  \bibinfo {author} {\bibfnamefont {G.}~\bibnamefont {Turnock}}, \ and\
  \bibinfo {author} {\bibfnamefont {B.~R.}\ \bibnamefont {Webber}},\ }\href
  {\doibase 10.1016/0370-2693(91)90196-W} {\bibfield  {journal} {\bibinfo
  {journal} {Phys. Lett. B}\ }\textbf {\bibinfo {volume} {269}},\ \bibinfo
  {pages} {432} (\bibinfo {year} {1991})}\BibitemShut {NoStop}%
\bibitem [{\citenamefont {Sjostrand}(1994)}]{Sjostrand:1994kzr}%
  \BibitemOpen
  \bibfield  {author} {\bibinfo {author} {\bibfnamefont {T.}~\bibnamefont
  {Sjostrand}},\ }\href {\doibase 10.1016/0010-4655(94)90132-5} {\bibfield
  {journal} {\bibinfo  {journal} {Comput. Phys. Commun.}\ }\textbf {\bibinfo
  {volume} {82}},\ \bibinfo {pages} {74} (\bibinfo {year} {1994})},\ \Eprint
  {http://arxiv.org/abs/hep-ph/9508391} {arXiv:hep-ph/9508391} \BibitemShut
  {NoStop}%
\bibitem [{\citenamefont {Dokshitzer}\ \emph {et~al.}(1988)\citenamefont
  {Dokshitzer}, \citenamefont {Khoze}, \citenamefont {Troian},\ and\
  \citenamefont {Mueller}}]{Dokshitzer:1987nm}%
  \BibitemOpen
  \bibfield  {author} {\bibinfo {author} {\bibfnamefont {Y.~L.}\ \bibnamefont
  {Dokshitzer}}, \bibinfo {author} {\bibfnamefont {V.~A.}\ \bibnamefont
  {Khoze}}, \bibinfo {author} {\bibfnamefont {S.~I.}\ \bibnamefont {Troian}}, \
  and\ \bibinfo {author} {\bibfnamefont {A.~H.}\ \bibnamefont {Mueller}},\
  }\href {\doibase 10.1103/RevModPhys.60.373} {\bibfield  {journal} {\bibinfo
  {journal} {Rev. Mod. Phys.}\ }\textbf {\bibinfo {volume} {60}},\ \bibinfo
  {pages} {373} (\bibinfo {year} {1988})}\BibitemShut {NoStop}%
\bibitem [{\citenamefont {Dokshitzer}\ \emph {et~al.}(1991)\citenamefont
  {Dokshitzer}, \citenamefont {Khoze}, \citenamefont {Mueller},\ and\
  \citenamefont {Troian}}]{Dokshitzer:1991wu}%
  \BibitemOpen
  \bibfield  {author} {\bibinfo {author} {\bibfnamefont {Y.~L.}\ \bibnamefont
  {Dokshitzer}}, \bibinfo {author} {\bibfnamefont {V.~A.}\ \bibnamefont
  {Khoze}}, \bibinfo {author} {\bibfnamefont {A.~H.}\ \bibnamefont {Mueller}},
  \ and\ \bibinfo {author} {\bibfnamefont {S.~I.}\ \bibnamefont {Troian}},\
  }\href@noop {} {\emph {\bibinfo {title} {{Basics of perturbative QCD}}}}\
  (\bibinfo  {publisher} {{Editions Frontieres}},\ \bibinfo {address} {Paris},\
  \bibinfo {year} {1991})\BibitemShut {NoStop}%
\bibitem [{\citenamefont {Aidala}\ \emph {et~al.}(2025)\citenamefont {Aidala},
  \citenamefont {Loomis}, \citenamefont {Martinez}, \citenamefont {Sassot},\
  and\ \citenamefont {Stratmann}}]{Aidala:2025kep}%
  \BibitemOpen
  \bibfield  {author} {\bibinfo {author} {\bibfnamefont {C.~A.}\ \bibnamefont
  {Aidala}}, \bibinfo {author} {\bibfnamefont {D.~A.}\ \bibnamefont {Loomis}},
  \bibinfo {author} {\bibfnamefont {R.~T.}\ \bibnamefont {Martinez}}, \bibinfo
  {author} {\bibfnamefont {R.}~\bibnamefont {Sassot}}, \ and\ \bibinfo {author}
  {\bibfnamefont {M.}~\bibnamefont {Stratmann}},\ }\href@noop {} {\  (\bibinfo
  {year} {2025})},\ \Eprint {http://arxiv.org/abs/2507.04887} {arXiv:2507.04887
  [hep-ph]} \BibitemShut {NoStop}%
\bibitem [{\citenamefont {Chen}\ \emph {et~al.}(2022)\citenamefont {Chen} \emph
  {et~al.}}]{Electron-PositronAlliance:2021kig}%
  \BibitemOpen
  \bibfield  {author} {\bibinfo {author} {\bibfnamefont {Y.}~\bibnamefont
  {Chen}} \emph {et~al.} (\bibinfo {collaboration} {Electron-Positron
  Alliance}),\ }\href {\doibase 10.1007/JHEP06(2022)008} {\bibfield  {journal}
  {\bibinfo  {journal} {JHEP}\ }\textbf {\bibinfo {volume} {06}},\ \bibinfo
  {pages} {008} (\bibinfo {year} {2022})},\ \Eprint
  {http://arxiv.org/abs/2111.09914} {arXiv:2111.09914 [hep-ex]} \BibitemShut
  {NoStop}%
\bibitem [{\citenamefont {Badea}\ \emph {et~al.}(2025)\citenamefont {Badea}
  \emph {et~al.}}]{Electron-PositronAlliance:2025hze}%
  \BibitemOpen
  \bibfield  {author} {\bibinfo {author} {\bibfnamefont {A.}~\bibnamefont
  {Badea}} \emph {et~al.} (\bibinfo {collaboration} {Electron-Positron
  Alliance}),\ }\href@noop {} {\  (\bibinfo {year} {2025})},\ \Eprint
  {http://arxiv.org/abs/2510.22038} {arXiv:2510.22038 [hep-ex]} \BibitemShut
  {NoStop}%
\bibitem [{\citenamefont {Bonino}\ \emph {et~al.}(2026)\citenamefont {Bonino},
  \citenamefont {Gehrmann-De~Ridder}, \citenamefont {Gehrmann}, \citenamefont
  {Huss}, \citenamefont {Merlotti},\ and\ \citenamefont
  {Stagnitto}}]{bonino_2026_18683133}%
  \BibitemOpen
  \bibfield  {author} {\bibinfo {author} {\bibfnamefont {L.}~\bibnamefont
  {Bonino}}, \bibinfo {author} {\bibfnamefont {A.}~\bibnamefont
  {Gehrmann-De~Ridder}}, \bibinfo {author} {\bibfnamefont {T.}~\bibnamefont
  {Gehrmann}}, \bibinfo {author} {\bibfnamefont {A.}~\bibnamefont {Huss}},
  \bibinfo {author} {\bibfnamefont {F.}~\bibnamefont {Merlotti}}, \ and\
  \bibinfo {author} {\bibfnamefont {G.}~\bibnamefont {Stagnitto}},\ }\href
  {\doibase 10.5281/zenodo.18683133} {\enquote {\bibinfo {title} {Precise qcd
  predictions for hadron-in-jet production in $e^+e^-$ collisions},}\ }
  (\bibinfo {year} {2026})\BibitemShut {NoStop}%
\end{thebibliography}%

\end{document}